\documentclass[]{IEEEphot}

\usepackage{amsmath,amssymb}
\usepackage{graphicx}
\usepackage{algorithm}
\usepackage{algpseudocode}

\begin{document}

\title{MIMO Codes for Uniform Illumination across Space and Time in VLC with Dimming Control}

\author{T. Uday, Abhinav Kumar,~\IEEEmembership{Member,~IEEE,}
       and L. Natarajan,~\IEEEmembership{Member,~IEEE}}

\affil{Department of Electrical Engineering, 
Indian Institute of Technology Hyderabad, Telangana, 502285 India.}  


\maketitle



\begin{abstract}
Indoor visible light communications (VLC) require simultaneous illumination and communication. Hence, uniformity in the illumination is a key consideration for user comfort and data transfer in VLC systems. Several run-length limited codes have been proposed in the literature which mitigate flicker for single input and single output VLC systems. 
Recently, codes have been proposed for multiple input multiple output (MIMO) VLC systems. However, uniform illumination along with dimming control has not been considered for MIMO VLC systems. Hence, in this paper, we propose codes with generalized algorithms for encoding and decoding that maintain consistency in the illumination across space and time and achieve the desired dimming level.
We present the expressions for code rate, run-length, and Hamming distance for the proposed codes.
Through Monte Carlo simulations, we compare the codeword error rate (CER) performance of the proposed codes
with zero-forcing, minimum mean square error, and maximum likelihood detectors for various number of transmit and receive antennas. We also compare the mutual information of the proposed codes for various number of transmit and receive antennas. The numerical results show that the proposed codes exhibit good performance while satisfying the design criteria. 
\end{abstract}


\section{Introduction}
Nowadays, light emitting diodes (LEDs) are extensively used for indoor illumination due to their several advantages over the conventional sources of light \cite{1}.
Visible light communications (VLC) uses intensity modulation of optical sources like LEDs for wireless data transfer to achieve high data rates with little effect on perceived illumination by humans \cite{1a,1b}. Given that the same LEDs are simultaneously used for illumination and communication in VLC, the user should not perceive changes in the intensity levels while the data is being transferred.
The binary stream of data to be transferred can contain large streams of zeros (assuming ON-OFF keying (OOK)) which causes flickering  whenever the duration of zeros exceeds the maximum flickering time period (MFTP) of 5 ms \cite{2}, and hence, can cause discomfort to the human eyes. To mitigate this flicker effect, several run-length limited (RLL) codes have been proposed for single input single output (SISO) VLC systems \cite{3,4,5,5a,5A1}. All codewords of the RLL codes may not always have equal weight \cite{3} which can cause inconsistency in the illumination. Hence, direct current (DC) balanced codes like 4B6B code \cite{5} have been proposed for SISO VLC systems. The DC balanced codes have equal number of zeros and ones in all of their codewords ensuring constant weight codewords.
A generalized space shift keying technique for multiple input multiple out (MIMO) VLC systems has been proposed in \cite{6,7,8}. However, consistency in illumination for all the transmit antennas (LEDs) has not been considered. In \cite{9}, a trace-orthogonal based 
 pulse position modulation space time block codes (STBC) has been designed that ensures constant illumination in a time slot. Similarly, novel angle diversity techniques for indoor MIMO VLC systems have been proposed in \cite{9A}. Optimal constellation design for $2 \times 2$ MIMO VLC system has been presented in \cite{9A1} that results in improved error performance over the existing methods. However, consistency in illumination across different time slots has not been considered in these works.
  
Given different times of the day, different brightness levels are required in order to maintain the desired ambiance \cite{9a}. To achieve desired ambiance, dimming levels have to be controlled. For this, addition of compensation symbols at the end of the RLL codes has been proposed for VLC systems in \cite{10}. Addition of the compensation symbols corresponding to the desired dimming level to improve the bit error rate performance has been proposed in \cite{10_self}. Novel schemes to achieve desired dimming levels for SISO VLC systems have been proposed in \cite{10_a,10_b}. However, to the best of our knowledge this paper is the first work that jointly considers uniform illumination across space and time along with dimming control for MIMO VLC systems.

The contributions of this work are as follows.
We propose novel code design that can be used to generate codes for MIMO VLC systems that provides uniform illumination across space and time conditioned on MFTP and maintain consistency in the illumination. Further, we achieve the desired dimming levels based on the number of transmit antennas. We consider the code rate, run-length, symbol error rate (SER), mutual information, and Hamming distance as the performance metrics for the proposed codes. We present simulation results comparing the performance of the proposed codes
with zero-forcing (ZF), minimum mean square error (MMSE), and maximum likelihood (ML) detectors for various number of transmit and receive antennas.
\begin{figure*}
\centering
  \centering
  \includegraphics[width=1.0\linewidth]{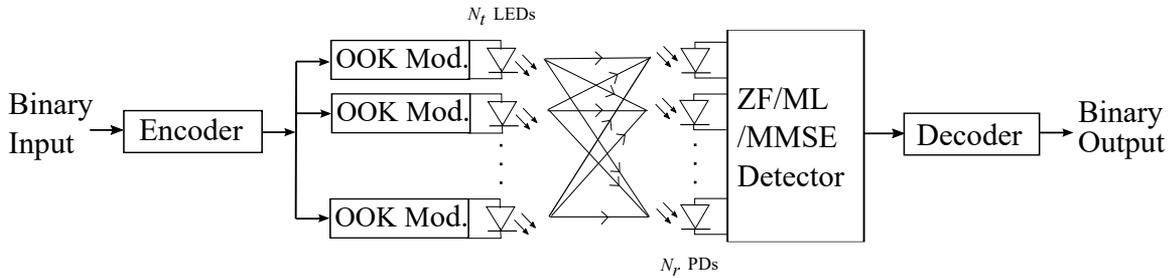}
  \caption{Schematic of the VLC system considered.}
  \label{Schematic}
\end{figure*}

The organization of the paper is as follows. In Section II, proposed system model is presented. In Section III, the algorithm for encoding and decoding of the proposed codes is discussed. Dimming support of the proposed codes with an algorithm to generate the code matrices that achieve the desired dimming level is presented in Section IV. In Section V, code rate, run-length, CER, mutual information, and Hamming distance are analyzed for the proposed codes. Numerical results are presented in Section VI. In Section VII, some concluding remarks along with the future work are discussed.

\section{System model}
The schematic of the VLC system considered in this work is shown in Fig.~\ref{Schematic}. We consider a MIMO VLC system with $N_t$ transmit antennas and $N_r$ receive antennas.
From the input bit stream at the transmitter (Tx), we assume that at a time $k$ bit length message vector is passed through the encoder. The output of the encoder is considered to be $N_t$ bits per time slot each of which is uniquely mapped to a Tx antenna. These bits are then OOK modulated and transmitted through the respective Tx antennas. At receiver (Rx), the received signals from the $N_r$ Rx antennas over multiple time slots are jointly passed through the detector as shown in Fig.~\ref{Schematic}. These equalized signals are processed by the decoder that generates a binary message vector of length $k$ at a time at the Rx.
Let $\mathbf{\underline{H}}$ denote the channel gain matrix given by
\[
\mathbf{\underline{H}} = \begin{bmatrix} 
    h_{1\,1} & h_{1\,2} & h_{1\,3} &\dots  &h_{1\,N_t} \\
    h_{2\,1} & h_{2\,2} & h_{2\,3} &\dots  &h_{2\,N_t} \\
    \vdots & \vdots & \vdots  &\ddots & \vdots \\
 h_{N_r\,1} & h_{N_r\,2} & h_{N_r\,3} &\dots  &h_{N_r\,N_t} \\
    \end{bmatrix}
,\]
where, $h_{j\, i}$ is the channel gain between $j^{th}$ Rx antenna and $i^{th}$ Tx antenna.
Considering additive white Gaussian noise (AWGN) channel as in \cite{10B}, the received signal, $\mathbf{\underline{Y}}$ can be represented as,
\begin{equation*}
\mathbf{\underline{Y}}=E_s \mathbf{\underline{H}}\, \mathbf{\underline{X}}^a+\mathbf{\underline{N}},
\end{equation*}
where, $E_s$ is the intensity multiplication factor of $\mathbf{\underline{X}}^a$, $\mathbf{\underline{X}}^a$ is $a^{th}$ the code matrix and $\mathbf{\underline{N}}$ is the independent and identically distributed (i.i.d) additive white Gaussian noise (AWGN) vector with 0 mean and co-variance matrix $\mathbf{\underline{\Sigma}}$.
We consider the following VLC channel in this work. However, the proposed codes are also valid for other VLC channel models.
\begin{figure}
\centering
  \centering
  \includegraphics[width=0.3\linewidth]{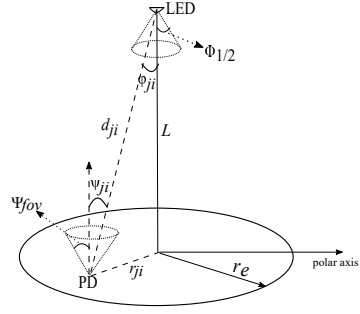}
  \caption{Figure showing Tx antenna and Rx antenna parameters.}
  \label{Schematic3}
\end{figure}

\subsection{VLC channel}
The DC channel gain \cite{10A} between $j^{th}$ Rx antenna and $i^{th}$ Tx antenna is given as
\begin{equation}
h_{j\, i}=\frac{(m+1)AR_p}{2\pi d_{j\, i}^2} \mathrm{cos}(\phi_{j\, i})^m T(\psi_{j\, i})g(\psi_{j\, i})\mathrm{cos}(\psi_{j\, i})\,,
\label{vlc_channel}
\end{equation}
where, $m$ is the order of Lambertian radiation pattern given by $m=-1/\mathrm{log}_2(\mathrm{cos}(\Phi_{1/2}))$  such that $\Phi_{1/2}$ is the angle at half power of LED, $A$ denotes the detection area of the photo detector (PD) at the Rx, $R_p$ denotes the responsivity of the PD, $T(\psi_{j\, i})$ represents the gain of the optical filter used at the Rx, $d_{j\, i}$ is the distance between the $j^{th}$ Rx antenna and the $i^{th}$ Tx antenna, $\psi_{j\, i}$ is the angle as shown in Fig.~\ref{Schematic3}, and $g(\psi_{j\, i})$ represents the gain of the optical concentrator given by

$$
g(\psi_{j\, i})=
\begin{cases}
\frac{\eta^2}{\mathrm{sin}(\psi_{fov})} &, 0\leq \psi_{j\, i}\leq \psi_{fov}\,,\\
0&, \psi_{j\, i}>\psi_{fov}\,,\\
\end{cases}
$$
where, $\eta$ is the reflective index of the optical concentrator at the Rx and $\psi_{fov}$ is the field of view of the photo detector at the Rx as shown in Fig.~\ref{Schematic3}. Note that, just for ease of illustration, we have represented the parameters for the $j^{th}$ Rx antenna and $i^{th}$ Tx antenna in Fig.~\ref{Schematic3}, the same can be generalized to any Rx antenna-Tx antenna pair in the considered VLC system.

Considering uniform movement of the Rx in the radial direction, as in \cite{10A}, the final expression for channel gain between $j^{th}$ Rx antenna and $i^{th}$ Tx antenna is given by
\begin{equation}
h_{ji}=\frac{C(m+1)L^{m+1}}{\left(r_{j\, i}^2+L^2 \right)^{\frac{m+3}{2}}}\,,
\label{rerep}
\end{equation}
where, $L$ is the vertical distance between $i^{th}$ LED and the surface, $r_{j\, i}$ is the radial distance between the $j^{th}$ Rx antenna and $i^{th}$ Tx antenna from the top view as shown in Fig.~\ref{Schematic3}, and 
$C=\frac{1}{2\pi}AR_pT(\psi_{j\, i})g(\psi_{j\, i})$. Note that \eqref{vlc_channel} is equivalent to \eqref{rerep}. However, \eqref{rerep} allows modeling of random movement of Rx in the Rx plane. Next, we present the proposed codes with generalized algorithms.
\section{Proposed codes}
Given $k$ length message vector and $N_t$ Tx antennas, we propose a method to map these $k$ bits to $N_t\times N_t$ code matrix.
We denote a $N_t\times N_t\,a^{th}$code matrix by $\mathbf{\underline{X}}^a$, where, $a\, \in \,\{0,1,\ldots,2^k-1\}$ such that
\[
\mathbf{\underline{X}}^a = \begin{bmatrix} 
    x_{1\,1}^a & x_{1\,2}^a & x_{1\,3}^a &\dots &x_{1\,{{N_t-1}}}^a &x_{1\,{N_t}}^a \\
    x_{2\,1}^a & x_{2\,2}^a & x_{2\,3}^a &\dots &x_{2\,{N_t-1}}^a &x_{2\,{N_t}}^a \\
    \vdots & \vdots & \vdots           &  \ddots     &\vdots    &\vdots \\
     x_{{N_t-1}\,1}^a & x_{{N_t-1}\,2}^a & x_{{N_t-1}\,3}^a &\dots &x_{{N_t-1}\,{N_t-1}}^a &x_{{N_t-1}\,{N_t}}^a \\
   x_{{N_t}\,1}^a & x_{{N_t}\,2}^a & x_{{N_t}\,3}^a &\dots &x_{{N_t}\,{N_t-1}}^a &x_{{N_t}\,{N_t}}^a \\
    \end{bmatrix}
,\]
where, $x_{j\,s}^a$ denotes the bit at $j^{th}$ Tx antenna in the $s^{th}$ time slot of the $a^{th}$ code matrix. We first propose codes for minimum possible dimming factor denoted by $\gamma$ and then generalize them for various achievable dimming factors, for a given $N_t$. Traditionally, in single Tx antenna VLC systems, as in \cite{10_self}, the dimming factor has been defined as the proportion of time the Tx antenna is ON. However, in MIMO VLC systems the dimming factor can be defined as the proportion of time each Tx antenna is ON while transmitting a code matrix. Hence, we define $\gamma$ as
\begin{equation}
\gamma=\frac{\mbox{ Total number of 1s in the code matrix}}{\mbox{ Total number of bits in the code matrix}}.
\label{dimm_def}
\end{equation}
To ensure constant dimming level, the $\gamma$ should not change between code matrices. To ensure uniformity in illumination per Tx antenna, each of the Tx antenna should be ON for same proportion of time. Thus, each row of the code matrix should have the same weight for a given $N_t$. Similarly, to ensure uniform illumination by the $N_t$ Tx antennas in a time slot, each column of the code matrix should have the same weight. Thus, the requirement of uniform illumination and constant dimming level across time slots and Tx antennas results in the following two constraints
\begin{eqnarray}
\sum_{s=1}^{N_t} x_{s\,j}^a = M \,,   \forall \,j\; \epsilon \{1,2,\ldots, N_t\},\label{con_0}\\
\label{con_1}
\sum_{j=1}^{N_t} x_{s\,j}^a = M\,,    \forall \,s\; \epsilon \{1,2,\ldots, N_t\},
\end{eqnarray}
where, $M$ should be equal to $\gamma N_t$. The minimum possible value of $M$ is 1. Hence, using \eqref{dimm_def}, \eqref{con_0}, and \eqref{con_1}, minimum possible value of $\gamma$ that can be achieved with uniform illumination and constant dimming is 1/$N_t$.

\begin{figure}
\begin{center}
  \includegraphics[width=0.3\columnwidth]{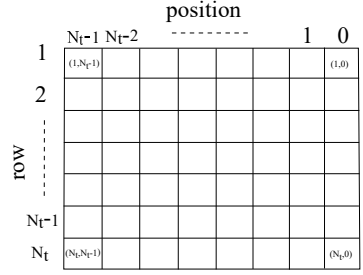}
  \caption{Format of the code matrix showing row number and position of 1s.}
  \label{codematrix}
  \end{center}
\end{figure}

We first consider the code design for $M=1$, i.e., $\gamma=1/N_t$. We will generalize this code design for various possible dimming factors in the next section. 
The proposed algorithm maps every $k$ bits for a given $N_t$ to a unique code matrix that has exactly one 1 in each row and each column. For example, a $N_t\times N_t$ identity matrix is a valid code matrix satisfying \eqref{con_0}, and \eqref{con_1}. A code matrix generated by this method will maintain uniformity in the illumination over space and time. The positions of 1s in the code matrix are decided by considering Fig.\ref{codematrix}. The column indices are considered from right to left and row indices are considered form top to bottom.
Consider the following lemma, where, the possibility for unique mapping of message vectors to code matrices is discussed.

\textit{\textbf{Lemma}} \textbf{1}. \textit{If $P_i=\left \lfloor R_i/(N_t-i)!\right \rfloor $, $1 \leq i \leq N_t$ and $R_i=R_{i-1} \mathrm{mod} (N_t-i+1)!$, $2 \leq i \leq N_t$, $0 \leq R_1 \leq 2^{N_t}-1$, then the set of $P_i$s denoted by $\{ P_i\}$ will differ in atleast one value for every distinct $R_1$, where $P_i$s and $R_i$s are integers.}

\textit{Proof}. See Appendix A for the proof.

\begin{algorithm}[t]
  \caption{Algorithm for encoding}\label{encoder}
  \begin{algorithmic}[1]
  \State INPUT: The number of Tx antennas, $N_t$.
  \State OUTPUT: $N_t$ x $N_t$ code matrix.
  \State Find the optimal value of message vector length, $k$ using constraint.
  \State From the binary stream of data each $k$ bits are considered at a time and they are converted to equivalent decimal values.
  \State Let $R_1$ be the integer initially corresponding to $k$ bits, $p_i$ be the position of 1 in the $i^{th}$ row, $R_{i}$ be the integer for the $i^{th}$ row, $p_{i}$ be the position of 1 in the $i^{th}$ row. 
  \State Initializations: Let val=$N_t$-1, $p_1$= $\left \lfloor{R_1/\;\text{val!}} \right \rfloor$.
      \For{i=2 to $N_t$}
        \State $R_{i}=R_{i-1}\;$ $\mathrm{mod}$ val!
        \State {val=val-1}
        \State $p_{i}=\left \lfloor{R_{i}/\;\text{val!}} \right \rfloor$
        \State  If there are 1s in the previous rows whose position is less than or equal to the current position, then increment the current $P_{i}$ by that many number of positions. During this increment process, if any position of 1s in the previous rows become less than or equal to the $P_{i}$ in the increment process, then the increment for that positions also should be considered.
           
    \State Update the code matrix here.   
    \EndFor
\State  Code matrix corresponding to the input $k$ length message vector is generated.

  \end{algorithmic}
\end{algorithm}

Let the code matrix be $\mathbf{\underline{X}}^a$, $R_1$ be the decimal equivalent of the $k$ binary bits, $R_{i}$ be the integer for the $i^{th}$ row and $P_{i}$ be the position of 1 in the $i^{th}$ row. Then for $1^{st}$ row the position of 1 is calculated as below
\begin{eqnarray*}
 (N_t-1).(N_t-2).(N_t-3)\ldots3.2.1.\; P_1 \leq R_1,\\
 (N_t-1)!P_1\leq R_1,\\
 P_1=\left \lfloor{R_1/(N_t-1)!} \right \rfloor,\\
     x_{1\, pos}^a= 
\begin{cases}
    1,& \text{if } pos = P_1\\
    0,              & \text{otherwise},\\
\end{cases}\\
\text{where,}\; pos \;\epsilon\;\{0,1,\ldots (n-1)\}.
\end{eqnarray*}

For any other row i.e., $i^{th}$ row and $i\leq\,N_t$, the position of 1s is given as
\begin{eqnarray*}
R_{i}=R_{i-1} \; \mathrm{mod} \; \left((N_t-i+1).(N_t-i).(N_t-i-1)\ldots 3.2.1\right),\\
R_{i}=R_{i-1}\; \mathrm{mod} \; (N_t-i+1)!,\\
(N_t-i).(N_t-i-1).(N_t-i-2)\ldots3.\;2.\;1.\; P_{i} \leq R_{i},\\
 (N_t-i)!P_{i}\leq R_{i},\\
 P_{i}=\left \lfloor{R_{i}/(N_t-i)!} \right \rfloor. \\
 \end{eqnarray*}
 If there are 1s in the previous rows whose position is less than or equal to the current position, then increment the current $P_{i}$ by that many number of positions. During this increment process, if any position of 1s in the previous rows become less than or equal to the $P_{i}$ in the increment process, then the increment for that positions also should be considered.
 \begin{eqnarray*}
     x_{i\,pos}^a= 
\begin{cases}
    1,& \text{if } pos = P_{i}\\
    0,              & \text{otherwise},
\end{cases}\\
\text{where,}\; i \;\epsilon\;\{2\ldots N_t\}.
\end{eqnarray*}

\begin{algorithm}[t]
  \caption{Algorithm for decoding}\label{decoding}
  \begin{algorithmic}[1]
  \State INPUT: Received code matrix, \underline{Y}.
  \State OUTPUT: Decoded binary output.
  \State Find the locations of 1 in the received vector after equalization. Let pos\_vec be the vector containing the position of 1s in the code matrix to be decoded.
  \State Initializations: pos\_vec1(1)=pos\_vec(1)  \% \textit{pos\_vec1 contains the actual position values}.
           \For{r=2 to $N_t$}
            \State var0=0
             \For{t=r-1 to 1 (decrease)}
                 \If {pos\_vec(r) $>$ pos\_vec(t)}
                    \State var0++
                 \EndIf
             \EndFor 
            \State pos\_vec1(r)=pos\_vec(r)-var0
           \EndFor
    \State dec\_val=0 and var1=$N_t$-1
       \For{r=1 to $N_t$}
                 \If {r $>$ 1}
                    \State dec\_val=dec\_val+var1! $\times$ pos\_vec1(r)
                    \Else
                    \State dec\_val=dec\_val+var1! $\times$ pos\_vec1(1)
                 \EndIf
                 \State var1-\,-
             \EndFor 
    \State Convert dec\_val to binary to get the message back.
  \end{algorithmic}
\end{algorithm}
%

The code matrices generated by Algorithm.\ref{encoder} are given in Appendix B for $N_t$ equal to 4.
Decoding the code matrix to get back the message is done post equalization of the received noisy version of the code matrix. Here we find the actual positions i.e., the values of $p_{i}$ ($i\,\geq\,2$) which are calculated at step 10 in Algorithm.\ref{encoder} by decreasing the position index as many times as the number of 1s on the right hand side of the current 1 in the rows above the current row in the code matrix. Since $1^{st}$ row does not have any rows above it, position of 1 can be found directly. These actual positions can be directly used to recover the decimal equivalent corresponding to binary message vector of length $k$ as shown in Algorithm.\ref{decoding}. (\textit{Note: If the binary equivalent has lesser length than $k$, append 0s on the left hand side till it's length becomes $k$.}). Next, we present the proposed algorithm for dimming control.

\section{Dimming control with the proposed codes}
The proposed code design can be used for dimming control also. 
The proposed codes can achieve any dimming  of the kind $f/N_t$ where $f\epsilon \left\{ 1,2, \ldots, N_t-1  \right\}$. To achieve dimming corresponding to  the dimming factor $1-(1/N_t)$, the code matrices generated by the proposed encoder can be directly complemented. To achieve any other dimming, consider the following algorithm
\begin{algorithm}
  \caption{Algorithm to achieve dimming}\label{dimmingAlgo}
  \begin{algorithmic}[1]
  \State INPUT: code matrix corresponding to $\gamma=1/N_t$ i.e., the output of Algorithm 1 and the other input required is desired $\gamma$ in the possible range which is $2/N_t\leq \gamma \leq (N_t-2)/N_t$.
  \State OUTPUT: code matrix corresponding the desired $\gamma$ per Tx antenna.
  \State Find the number of 1s that are required to achieve the given $\gamma$ excluding the one 1 which is already present. This number is given by $f=\gamma N_t-1$.
   \For{\texttt{r=1 to $N_t$}}
           \State Find the location of 1 in $r^{th}$ row of the code matrix taken.
           \State Replace $f$ 0s which are on the right hand side of the 1 with ones.
           \State If location of replacement exceeds the code matrix dimension, start replacing from first location.
   \EndFor 
  \end{algorithmic}
\end{algorithm}
The code matrices generated by Algorithm.\ref{dimmingAlgo} are given in Appendix B for $N_t$ equal to 4 and $\gamma$ equal to 0.5 and 0.75.
Let $Wt_{\gamma actual}(\mathbf{\underline{X}}^a)$ be the weight of the code matrix output of the proposed algorithm and $Wt_{\gamma desired}(\mathbf{\underline{X}}^a)$  be the weight of the code matrix which achieves desired $\gamma$ per Tx antenna.
 $Wt_{\gamma actual}(\mathbf{\underline{X}}^a)$ =$1/N_t$ and $Wt_{\gamma desired}(\mathbf{\underline{X}}^a)=\gamma N_t^2$
The the relation between $Wt_{\gamma actual}(\mathbf{\underline{X}}^a)$ and  $Wt_{\gamma desired}$ is given as 
\begin{equation}
  Wt_{\gamma desired}(\mathbf{\underline{X}}^a)=\gamma  \left(Wt_{\gamma actual}(\mathbf{\underline{X}}^a)\right) ^2.
\end{equation}
Since actual code matrices and their complemented version represents the lower and upper bounds on the achievable dimming for given $N_t$, few comparisons and relation corresponding to them are given below.

Let the code matrices generated by the proposed Algorithm.\ref{encoder} be the actual code matrices. Let $\gamma_{actual}$ be the dimming factor per antenna for the actual version and $\gamma_c$ be the dimming factor per antenna for the complemented version. Then,
$\gamma_{actual}=1/N_t$ and $\gamma_c=1-(1/N_t)$.
The weight of the code matrix for the complemented version of the code matrices let it be $Wt_c(\mathbf{\underline{X}}^a)$. Then the relation between $Wt_{\gamma actual}(\mathbf{\underline{X}}^a)$ and $Wt_c(\mathbf{\underline{X}}^a)$ can be given as follows
\begin{eqnarray*}
Wt_{\gamma actual}(\mathbf{\underline{X}}^a)=N_t,\\
Wt_c(\mathbf{\underline{X}}^a)=N_t^2-N_t,\\
Wt_c(\mathbf{\underline{X}}^a)=N_t(N_t-1),\\
Wt_c(\mathbf{\underline{X}}^a)=Wt_{\gamma actual}(\mathbf{\underline{X}}^a)(Wt_{\gamma actual}(\mathbf{\underline{X}}^a)-1).\\
\end{eqnarray*}

Consider Table.\ref{tableweights}, where, complemented and actual versions are compared for few values of $N_t$.

\begin{table}[t]
\begin{center}
		\caption{Complemented Vs actual codes}
		\label{tableweights}
\begin{tabular}{|c|c|c|c|c|c|c|}
\hline
 $N_t$ & $\gamma_{actual}$ & $\gamma_{c}$ & $Wt_{\gamma actual}(\mathbf{\underline{X}}^a)$ &$Wt_{c}(\mathbf{\underline{X}}^a)$  \\
 \hline
4 & 0.2500 & 0.750 & 4 & 12\\
5 & 0.2000 & 0.800 & 5 & 20\\
6 & 0.1667 & 0.833 & 6 & 30\\
7 & 0.1428 & 0.857 & 7 & 42\\
8 & 0.1250 & 0.875 & 8 & 56\\
 \hline
\end{tabular} 
\end{center}
\end{table}
where, $RL_{actual}$ and $RL_c$ be the run length per Tx antenna for the actual and complemented versions respectively.
Next, we present the performance metrics considered for analyzing the proposed codes.
 
\section{Performance analysis}
In this section the performance of the proposed codes is analyzed in terms of code matrix error rate, mutual information and minimum Hamming distance.
\subsection{Code rate}
\begin{figure}
\centering
  \centering
  \includegraphics[width=0.3\columnwidth]{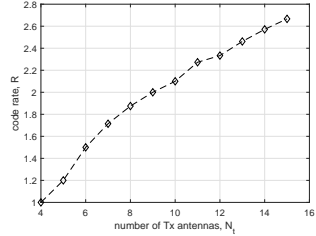}
  \caption{Variation of code rate with respect to number of Tx antennas, $N_t$ for the proposed codes.}
  \label{coderatevsnt}
\end{figure}

For $M=1$, from \eqref{con_1}, each Tx antenna is ON exactly for one time slot in a code matrix. Hence, for $N_t$ Tx antennas we have $N_t!$ combinations of code matrices that satisfy \eqref{con_0} and \eqref{con_1}. Given message vector length as $k$ bits, the total possible combinations of the message vectors is $2^k$. Each of the message vectors has to be uniquely mapped to a code matrix. Hence, we have the following constraint
\begin{equation}
 2^k \leq N_t!\,.
 \label{rate}
\end{equation}
The code rate is typically defined as the number of bits transmitted per time slot \cite{coderate}. Hence, for the proposed codes, the code rate denoted by $R$ will be equal to $k/N_t$. Thus, to maximize code rate, using \eqref{rate}, we have $k=\lfloor{\mathrm{log}_2(N_t!)}\rfloor$ and 
\begin{equation*}
    R=\frac{\left \lfloor{\mathrm{log}_2(N_t!)}\right \rfloor}{N_t}.
\end{equation*}
Note that for the proposed codes, the code rate increases with $N_t$ as shown in Fig.~\ref{coderatevsnt}.

%
\subsection{Run-length}
The run-length of the proposed codes is defined per Tx antenna. 
In VLC context, run-length is defined as the number of continuous 0s in the binary stream of data. Here, for MIMO codes, run-length is defined per Tx antenna.
For the actual version of the codes i.e., codes generated by Algorithm.\ref{encoder}, maximum run-length is 2$N_t$-2 which can be seen from the format of the code matrices and for the complemented version of the codes, the maximum run-length is always 2 irrespective of the value of $N_t$. Consider the following example code matrices for different values of $\gamma$ for the proposed codes for $N_t$=4.

\textit{Example:}

\[
\mathbf{\underline{X}}^a(\gamma=0.25)=\begin{bmatrix}
    0  &  0 & 0 & 1     \\
    0  &  0 & 1 & 0     \\ 
    0  &  1 & 0 & 0     \\
    1  &  0 & 0 & 0     \\    
\end{bmatrix},\]
\[
 \mathbf{\underline{X}}^a(\gamma=0.50)=
\begin{bmatrix}
    1  &  0 & 0 & 1     \\
    0  &  0 & 1 & 1     \\ 
    0  &  1 & 1 & 0     \\
    1  &  1 & 0 & 0     \\
\end{bmatrix}
,\]
\[
 \mathbf{\underline{X}}^a(\gamma=0.75)=
\begin{bmatrix}
    1  &  1 & 0 & 1     \\
    1  &  0 & 1 & 1     \\ 
    0  &  1 & 1 & 1     \\
    1  &  1 & 1 & 0     \\
\end{bmatrix}  
.\]

$\mathbf{\underline{X}}^a(\gamma=0.25)$ can give a maximum run-length of 6 per Tx antenna if the next transmitted code matrix has 0001 for $1^{st}$ Tx antenna and accordingly for the other Tx antennas.
$\mathbf{\underline{X}}^a(\gamma=0.75)$ can give a maximum run-length of 2 per Tx antenna if the next transmitted code matrix has 0111 for $4^{th}$ Tx antenna and accordingly for the other Tx antennas.
Similarly, for given $N_t$ and from Algorithm.\ref{dimmingAlgo} which helps in achieving desired $\gamma$, the run-length for any desired and possible $\gamma$ is given as
\begin{align}\nonumber
RL&=2N_t-2\gamma N_t\\
&=2N_t(1-\gamma).
\label{run-len_dimm}
\end{align}

Maximum possible run-length occurs when the dimming level is least for a given $N_t$. Consider Table.\ref{boundonnt}, where, the value of $N_t$ is calculated for few values of $T_{b}$.

A flicker is observed in a VLC system whenever the duration of 0s exceeds the time corresponding to MFTP. The MFTP is defined as
the maximum time period over which the change in light intensity is not perceived by the human eye.
The MFTP for human eye is 5ms or frequency less than 200Hz \cite{2}.
For the proposed codes, given $N_t$ number of Tx antennas, maximum run-length is $N_t-1$ per Tx antenna in a code matrix and hence, $2(N_t-1)$ in the data stream transmitted by any Tx antenna. Maximum run-length for given $N_t$ occurs for least possible $\gamma$ ($M$=1).

Let $T_{b}$ be the time period of a bit, either 0 or 1.  For given $N_t$ and for maximum possible run-length, to avoid flickering, the relation between MFTP and $T_{b}$ of the proposed codes is given as
\begin{equation}
    \left(2N_t-2\right)T_{b} < MFTP.
    \label{prebound}
\end{equation}
Then, from \eqref{prebound} the upper bound on the value of $N_t$ is given by the following equation.,
\begin{equation*}
    N_t=\left \lfloor{\frac{MFTP+2T_{b}}{2T_{b}}} \right \rfloor.
    \label{boundonn}
\end{equation*}
\begin{figure}
\centering
\begin{minipage}{.5\textwidth}
  \centering
  \includegraphics[width=.8\linewidth]{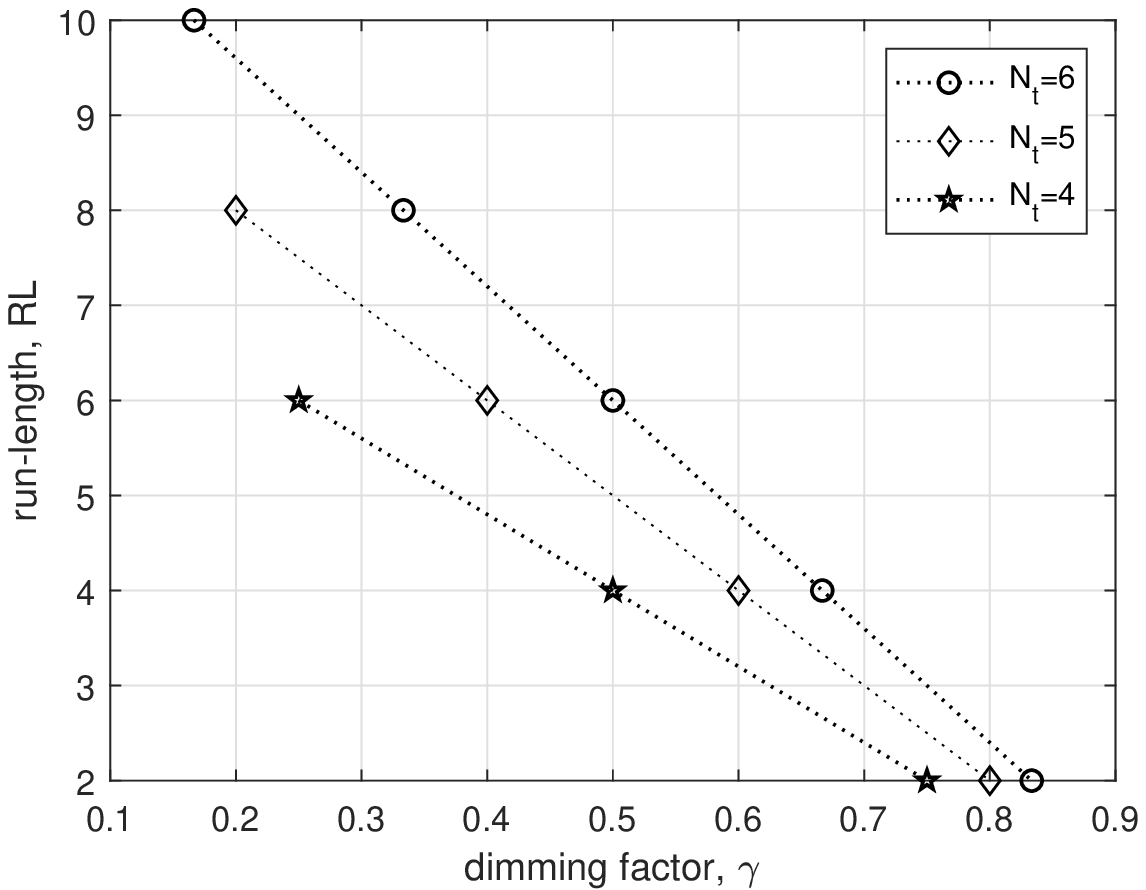}
  \caption{Run-length per Tx antenna by varying $\gamma$.}
  \label{dimming_RL}
\end{minipage}%
\begin{minipage}{.5\textwidth}
  \centering
  \includegraphics[width=.8\linewidth]{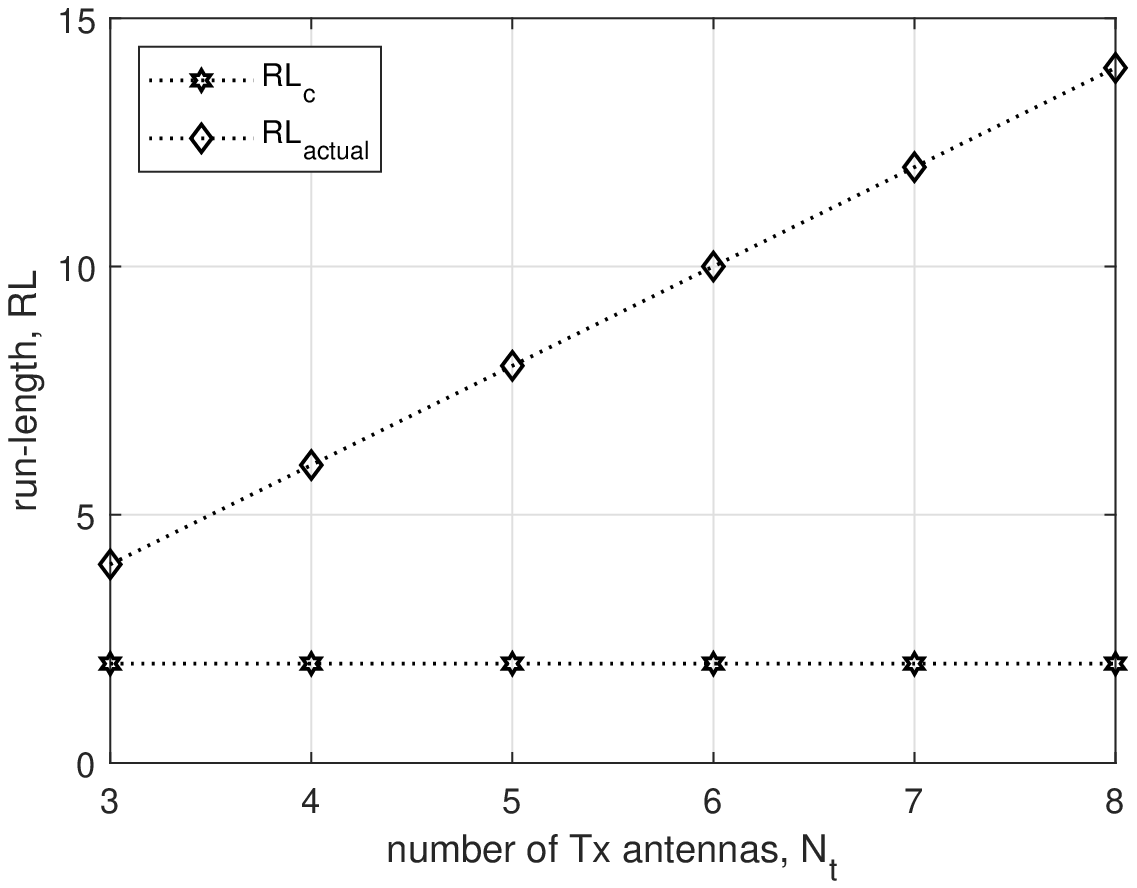}
  \caption{Run-length per Tx antenna for complemented and actual version of the proposed codes.}
  \label{RLVsN_t}
\end{minipage}
\end{figure}
In Fig.\ref{dimming_RL}, for given $N_t$, $RL$s are compared for all possible values of $\gamma$ using \eqref{run-len_dimm}. In Fig.\ref{RLVsN_t}, maximum possible $RL$ (for $\gamma=1/N_t$) for given $N_t$ are compared.

\begin{table}[t]
\begin{center}
		\caption{Upper bound on $N_t$ to achieve minimum possible dimming}
		\label{boundonnt}
\begin{tabular}{|c|c|}
 \hline
 $T_{b}$&$N_t$\\\hline
  0.2 ms & 13\\\hline
  0.1 ms & 26  \\\hline
  50 $\mu$s   & 51\\\hline
 20 $\mu$s   & 126\\\hline
  1 $\mu$s   & 2501\\\hline
\end{tabular} 
\end{center}
\end{table}
\subsection{Codeword error rate (CER)}
Union bound on CER is calculated using pairwise error probability (PEP) \cite{10C}. The expression for CER is given as
\begin{equation}
CER \leq \mathop{\mathbb{E}_\mathbf{\underline{H}}} \left( \frac{2}{2^k} \sum_{a=1}^{2^k-1}\sum_{b=a+1}^{2^k} PEP(\mathbf{\underline{X}}^a,\mathbf{\underline{X}}^b)\right),
\label{SER}
\end{equation} 
Where, $\mathop{\mathbb{E}_\mathbf{\underline{H}}}$ denotes expectation over channel gain matrix and $PEP(\mathbf{\underline{X}}^a,\mathbf{\underline{X}}^b)$ denotes $PEP$ between two code matrices $\mathbf{\underline{X}}^a$ and $\mathbf{\underline{X}}^b$ ($a\neq b$).
Then, the expression for $PEP$ can be calculated as shown below
\begin{equation*}
PEP(\mathbf{\underline{X}}^a, \mathbf{\underline{X}}^b)=Pr\left \{  ||\mathbf{\underline{Y}}-E_s\mathbf{\underline{H}}\,\mathbf{\underline{X}}^a||^2>||\mathbf{\underline{Y}}-E_s\mathbf{\underline{H}}\,\mathbf{\underline{X}}^b||^2 \bigg\rvert \mathbf{\underline{X}}^a    \right \}.
\end{equation*}
By substituting $\mathbf{\underline{Y}}=E_s\mathbf{\underline{H}}\,\mathbf{\underline{X}}^a+\mathbf{\underline{N}}$, the expression can be evaluated as
\begin{equation}
PEP(\mathbf{\underline{X}}^a,\mathbf{\underline{X}}^b)=Q\left(\frac{E_s||\mathbf{\underline{H}}(\mathbf{\underline{X}}^a-\mathbf{\underline{X}}^b)||}{\sqrt{2N_0}} \right).
\label{pep}
\end{equation}
By substituting \eqref{pep} in \eqref{SER}, the final expression in $Q(.)$ function can be given as follows
\begin{equation*}
CER\leq \frac{1}{2^{k-1}} \sum_{a=1}^{2^k-1}\sum_{b=a+1}^{2^k}\mathop{\mathbb{E}_\mathbf{\underline{H}}}\left( Q\left(E_s\sqrt{\frac{||\mathbf{\underline{H}}(\mathbf{\underline{X}}^a-\mathbf{\underline{X}}^b)||^2}{2N_0}} \right)\right).
\end{equation*}
In the above expression of CER, $||\mathbf{\underline{H}}(\mathbf{\underline{X}}^a-\mathbf{\underline{X}}^b)||^2$ is given by
\begin{equation*}
||\mathbf{\underline{H}}(\mathbf{\underline{X}}^a-\mathbf{\underline{X}}^b)||^2=\sum_{s=1}^{N_t}\sum_{j=1}^{N_r}\bigg\rvert\sum_{i=1}^{N_t}h_{j\,i}\left(x^a_{i\,s}-x^b_{i\,s} \right)\bigg\rvert^2,
\end{equation*}
where, $x^a_{i\,s}$ denotes value in $i^{th}$ row and the $s^{th}$ column of the $a^{th}$ code matrix and $Q(w)=\frac{1}{2\pi}\int_{w}^{\infty} \,exp(-u^2/2)du$.
\subsection{Mutual information}
Let $I(\mathbf{\underline{X}}^a;\mathbf{\underline{Y}}|\mathbf{\underline{H}})$ be the mutual information between $\mathbf{\underline{X}}^a$ and $\mathbf{\underline{Y}}$ given the channel gain matrix $\mathbf{\underline{H}}$. The number of channel uses is same as the number of Tx antennas from the code design. Then, 
 for $N_t$ channel uses, the mutual information \cite{10a} is given as
\begin{equation}
I(\mathbf{\underline{X}}^a;\mathbf{\underline{Y}}|\mathbf{\underline{H}})=\frac{1}{N_t}\left(h(\mathbf{\underline{Y}}|\mathbf{\underline{H}})-h(\mathbf{\underline{Y}}|\mathbf{\underline{H}},\mathbf{\underline{X}}^a)\right),
\label{mi_exp}
\end{equation}
where, $h(.)$ is the differential entropy function.
For given $\mathbf{\underline{H}},\, \mathbf{\underline{X}}^a$, $h(\mathbf{\underline{Y}}|\mathbf{\underline{H}},\mathbf{\underline{X}}^a)$ is given by the following equation., 
\begin{align*}\nonumber
h(\mathbf{\underline{Y}}|\mathbf{\underline{H}},\mathbf{\underline{X}}^a)&=h((E_s\mathbf{\underline{H}}\,\mathbf{\underline{X}}^a+\mathbf{\underline{N}})|\mathbf{\underline{H}},\mathbf{\underline{X}}^a),\\
&=h(\mathbf{\underline{N}}).
\end{align*}
Since $\mathbf{\underline{N}}$ has the dimensions $N_r\times N_t$ and AWGN is assumed, $h(\mathbf{\underline{N}})$ has to be calculated using matrix normal distribution.
Probability density function (pdf) of matrix normal distribution \cite{11,12} is given as
\begin{equation}
\mathcal{MN}_{(n,p)}=\frac{exp\left(\frac{1}{2} tr\left(\mathbf{\underline{V}}^{-1}(\mathbf{\underline{S}}-\mathbf{\underline{M}})^T\mathbf{\underline{U}}^{-1}(\mathbf{\underline{S}}-\mathbf{\underline{M}}) \right)\right)}{ (2\pi)^{np/2}|\mathbf{\underline{V}}|^{n/2}|\mathbf{\underline{U}}|^{p/2}},
\label{matmormaldist}
\end{equation}
where, $\mathbf{\underline{S}}$ is the Gaussian random matrix of dimensions $n\times p$, $\mathbf{\underline{M}}$ is the mean matrix of dimensions $n\times p$,
$\mathbf{\underline{V}}$ is the scale matrix of dimensions $n\times n$, and $\mathbf{\underline{U}}$ is the scale matrix of dimensions $p\times p$. From the assumption of i.i.d noise components, the expression for $h(\mathbf{\underline{N}})$ can be computed as below
\begin{equation}
h(\mathbf{\underline{N}})=\frac{1}{2} \mathrm{log}_2\left( (2\pi e)^{N_tN_r}|\mathbf{\underline{\Sigma}}|^{N_t}  \right). 
\label{hofn}
\end{equation}
Assuming uniform distribution for code matrices, the weight of each Gaussian in the Gaussian mixture is equal and hence, the final expression for mutual information using \eqref{mi_exp}, \eqref{matmormaldist} and \eqref{hofn} can be evaluated as below

\begin{eqnarray*}
I(\mathbf{\underline{X}}^a;\mathbf{\underline{Y}}|\mathbf{\underline{H}})=\frac{1}{N_t} \left( \mathop{\mathbb{E}} \left(\mathrm{log}_2\left(\frac{1}{f(y|\mathbf{\underline{H}})}\right)\right)-
\frac{1}{2}\mathrm{log}_2\left( (2\pi e)^{N_tN_r}|\mathbf{\underline{\Sigma}} |^{N_t}\right)\right),
\end{eqnarray*}
where, $\mathop{\mathbb{E}}$ is the expectation function and
\begin{equation*}
f(y|\mathbf{\underline{H}})=\frac{\sum_{\mathbf{\underline{X}}^a\epsilon\mathcal{X}}exp\left( -\frac{1}{2}tr\left((\mathbf{\underline{Y}}-E_s\mathbf{\underline{H}}\, \mathbf{\underline{X}}^a)^T \mathbf{\underline{\Sigma}}^{-1} (\mathbf{\underline{Y}}-E_s\mathbf{\underline{H}}\, \mathbf{\underline{X}}^a)    \right)\right)}{(2\pi)^{N_tN_r/2} 2^k |\mathbf{\underline{\Sigma}}|^{N_t/2}},
\end{equation*}
where, $tr(.)$ denotes the trace of a matrix, $f(.)$ is the probability density function of $y$ given $\mathbf{\underline{H}}$, and $\mathcal{X}$ is the set of all code matrices for given $N_t$ and $\gamma$.

\begin{figure}
\centering
\begin{minipage}{.5\textwidth}
  \centering
  \includegraphics[width=.8\linewidth]{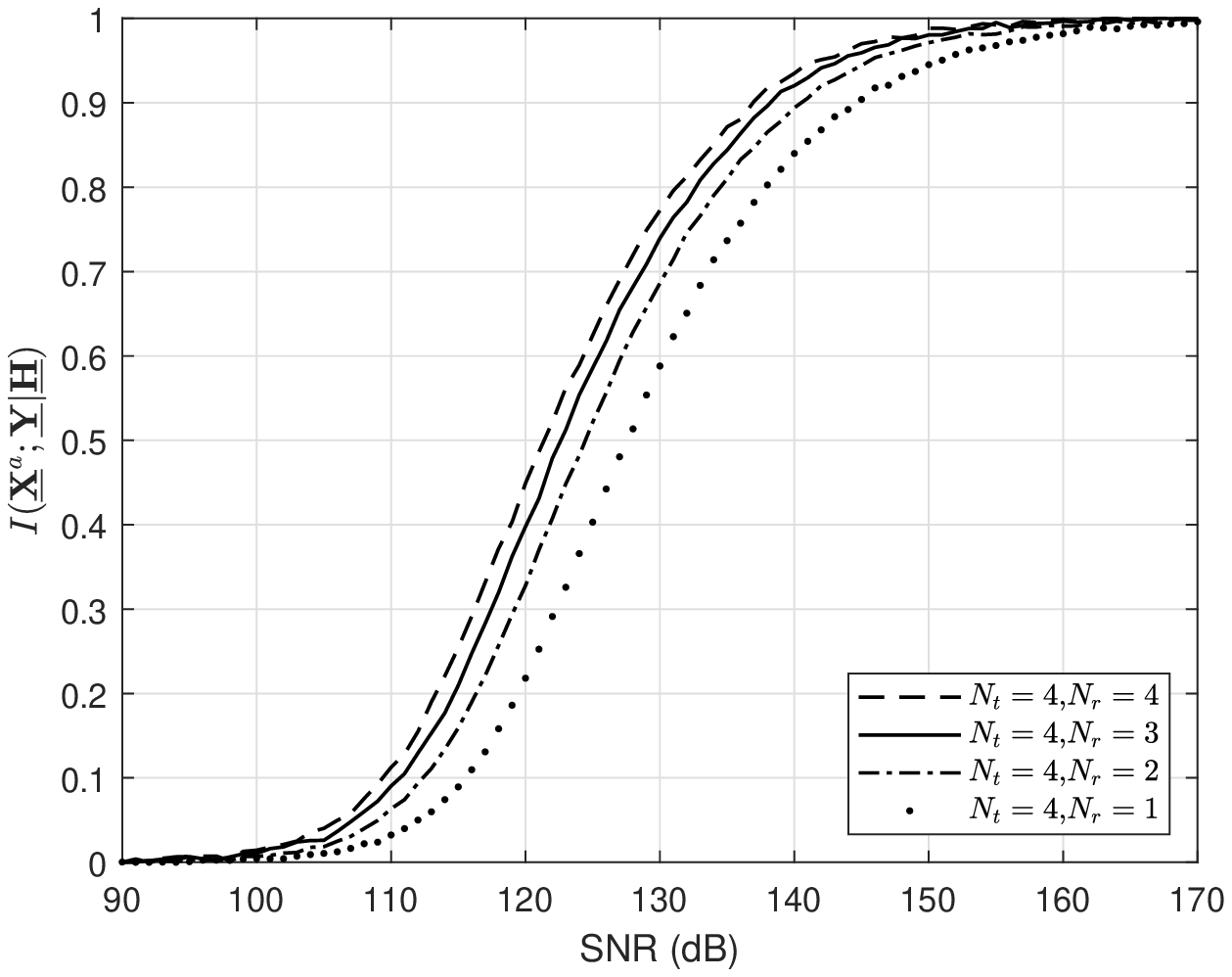}
  \caption{Mutual information plots for 4 Tx antennas and varying number of Rx antennas for $\gamma$=0.25.}
  \label{MI_1}
\end{minipage}%
\begin{minipage}{.5\textwidth}
  \centering
  \includegraphics[width=.8\linewidth]{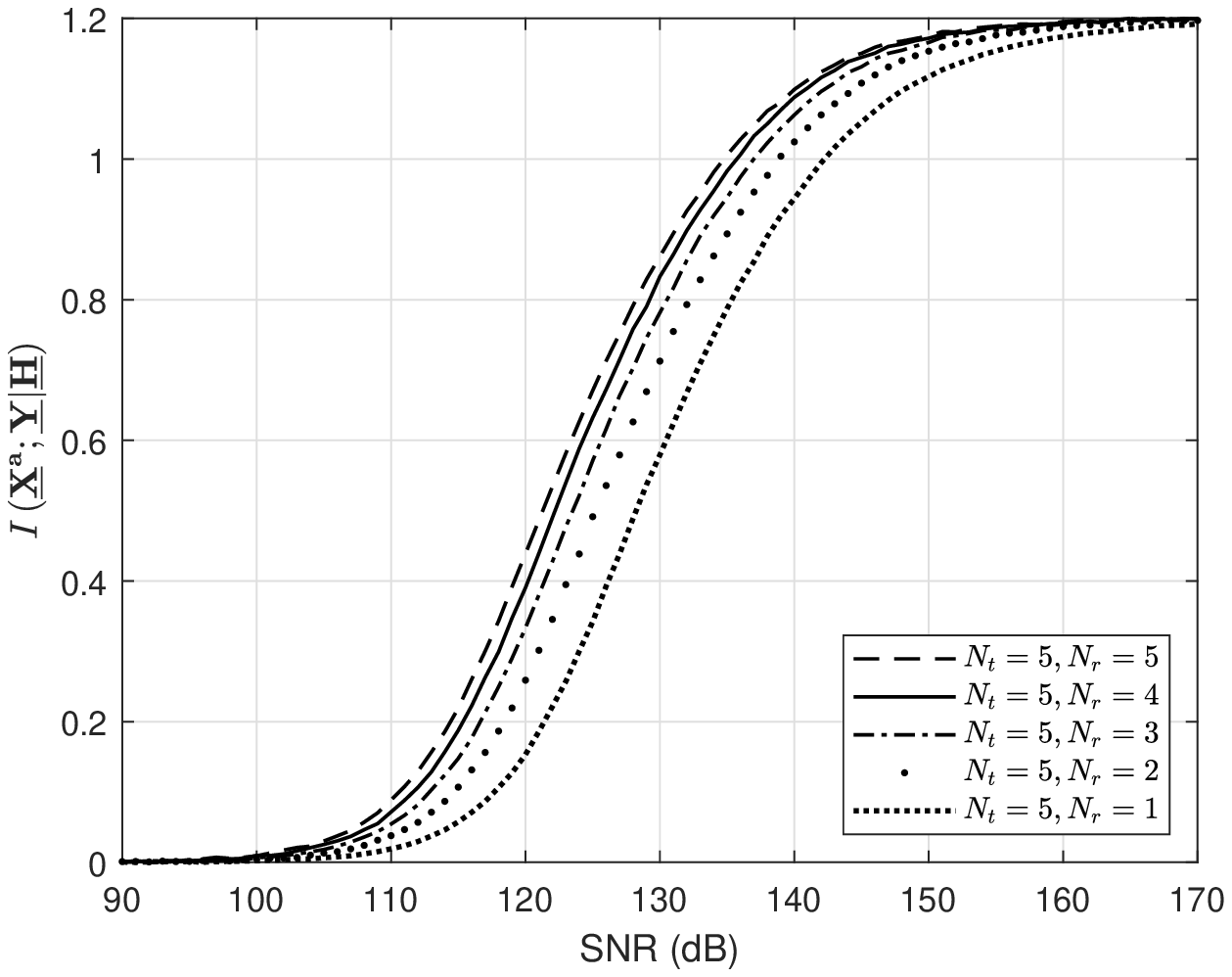}
  \caption{Mutual information plots for 5 Tx antennas and varying number of Rx antennas for $\gamma$=0.20.}
  \label{MI_2}
\end{minipage}
\end{figure}

\subsection{Minimum Hamming distance}
Since the output of the proposed encoder is binary code matrices, we consider minimum Hamming distance as the performance metric. However, minimum Euclidean distance can also be considered as the code matrices are intensity modulated before transmission.
Let $d_{min}$ be the minimum Hamming distance between any pair of code matrices $\mathbf{\underline{X}}^a$ and $\mathbf{\underline{X}}^b$ (a $\neq$ b). Let $\mathbf{\underline{X}}^a$ be denoted as

 \[
\mathbf{\underline{X}}^a = \begin{bmatrix} 
    \underline{X}_{1}^a  \\
    \underline{X}_{2}^a \\
    \underline{X}_{3}^a \\
    \vdots \\
 \underline{X}_{{N_t}}^a  
    \end{bmatrix}
,\]
where, $\underline{X}_{i}^a$ denotes the $i^{th}$ row of the $a^{th}$ code matrix.
 $d_{min}$ is first computed for $\gamma=1/N_t$ and then it is generalized for achievable arbitrary $\gamma$ whose range is conditioned on $N_t$. By the design of the codes each row  and  each column of any code matrix has only one 1 i.e., each code matrix is a permutation matrix. Let  $\mathbf{\underline{X}}^a$ be the first code matrix and $\mathbf{\underline{X}}^b$ be the second code matrix  which is formed by swapping exactly two rows of $\mathbf{\underline{X}}^a$. Let $l$ be the position of 1 in $\underline{X}_{i}^a$ and $l'$ be the position of 1 in $\underline{X}_{j}^a$ and $l\neq l'$. One such case can be shown as below
  \[
\mathbf{\underline{X}}^a = \begin{bmatrix} 
    \underline{X}_{1}^a  \\
    \underline{X}_{2}^a \\
    \underline{X}_{3}^a \\
    \vdots \\
 \underline{X}_{{N_t-1}}^a\\
 \underline{X}_{{N_t}}^a  
    \end{bmatrix},\,
    \mathbf{\underline{X}}^b = \begin{bmatrix} 
    \underline{X}_{1}^b  \\
    \underline{X}_{2}^b \\
    \underline{X}_{3}^b \\
    \vdots \\
 \underline{X}_{{N_t-1}}^b\\
 \underline{X}_{{N_t}}^b  
    \end{bmatrix}
,\]
where, $\underline{X}_{1}^b=\underline{X}_{1}^a$, $\underline{X}_{2}^b=\underline{X}_{2}^a$, $\ldots$, $\underline{X}_{{N_t-1}}^b=\underline{X}_{{N_t}}^a$, $\underline{X}_{N_t}^a=\underline{X}_{N_t-1}^a$. Now, let 
 $\underline{X}_{{N_t-1}}^a=[0\,0\,\ldots\,0\,0\,1\,0\,0\ldots\,0\,0]$, 
 $\underline{X}_{{N_t}}^a=[0\,0\,\ldots\,0\,0\,0\,1\,0\ldots\,0\,0]$. 
Then,
\begin{align*}
 d_{min}&=2\left(Wt(\underline{X}_{i}^a)+Wt(\underline{X}_{j}^a)\right),\\
        &=2(1+1),\\
        &=4.
\end{align*}
Similarly, for arbitrary $\gamma$ say $\gamma=q/N_t$ where $2\leq q\leq N_t-1$ and from Step 6 and 7 of Algorithm.\ref{dimmingAlgo} which says about the position of 1s to achieve desired $\gamma$. Consider the following example. Let
$\underline{X}_{i}^a=[0\,0\,\ldots\,0\,0\,1\,1\,1\ldots\,1\,1\,0\,0\ldots\,0$] ,
$\underline{X}_{j}^a=[0\,0\,\ldots\,0\,0\,0\,1\,1\ldots\,1\,1\,1\,0\ldots\,0$].
Then, $d_{min}$ can be given as below 
\begin{align*}
 d_{min}&=2\left(Wt(\underline{X}_{i}^a)+Wt(\underline{X}_{j}^a)\right)
 -4(\mbox{number of 1s common to both the rows}),\\
      &=2\left(Wt(\underline{X}_{i}^a)+Wt(\underline{X}_{j}^a)\right)-4(\gamma N_t-1),\\
       &=2\left(\gamma N_t+\gamma N_t\right)-4(\gamma N_t-1),\\
       &=4\gamma N_t-4\gamma N_t+4,\\
       &=4.
\end{align*}
Therefore, the $d_{min}$ for the proposed codes is 4. Similarly, minimum Euclidean distance for the proposed codes is 2$\sqrt{E_s}$. Next, we present numerical results comparing various cases of proposed codes.     
\section{Numerical results}

\begin{figure}
\centering
  \centering
  \includegraphics[width=0.3\linewidth]{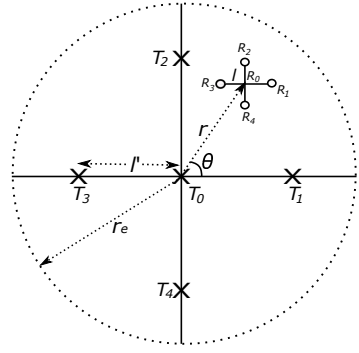}
  \caption{Top view of the scenario considered.}
  \label{Schematic2}
\end{figure}

 Assuming circular area of coverage, the top view of the system can be shown as in Fig.~\ref{Schematic2}. All the Tx antennas are assumed to be placed at equal distance from the center of the Tx, $T_0$. The Rx is assumed to move uniformly in the coverage area with the Rx plane horizontal to the surface and with constant orientation as shown in Fig.~\ref{Schematic2}. Let center of the surface be (0,0) i.e., the position of $T_0$ from the top view. Let $T_i$ denote the $i^{th}$ Tx antenna and $R_i$ denote the $i^{th}$ Rx antenna. The distance between $T_0$ and $T_i$ for $i\geq 1 $ be $l'$ and the distance between $R_0$ and $R_i$ for $i\geq 1 $ be $l$.
For 4 Tx antennas case, the Tx antennas assumed are {$T_i$} for $1\leq i\leq 4$. This is done just to maintain the symmetry in the VLC system. For the same case with $a$ Rx antennas ($a\, \leq\, 4)$, they are {$R_i$} for $1\leq i\leq a$. For 5 Tx antennas case the $5^{th}$ Tx antenna is taken to be $T_0$ and for $5^{th}$ Rx antenna, it is taken at $R_0$. Let $(c_i,d_i)$ be the position of $R_i$. In the proposed work, the reference point for the Rx is taken to be $R_0$ whose co-ordinates are calculated from the independent and uniformly distributed $r$ and $\theta$ which are shown in Fig.~\ref{Schematic2}. From the above considerations the position of any Rx antenna can be calculated using simple trigonometry analysis as given below. 

The position of $R_0$ is given as $(c_0,d_0)$ where $c_0=r\mathrm{cos}\theta$ and $d_0=r\mathrm{sin}\theta$. Then $c_1=c_0+l$ and $d_1=d_0$ which gives the position of $R_1$. $c_2=c_0$ and $d_2=d_0+l$ which gives the position of $R_2$. $c_3=c_0-l$ and $d_3=d_0$ which gives the position of $R_3$. $c_4=c_0$ and $d_4=d_0-l$ which gives the position of $R_4$. Note that the boundary for $R_0$ is taken to be $r_e$, but some Rx antenna can be outside the boundary.

\begin{figure}
\centering
\begin{minipage}{.5\textwidth}
  \centering
  \includegraphics[width=.8\linewidth]{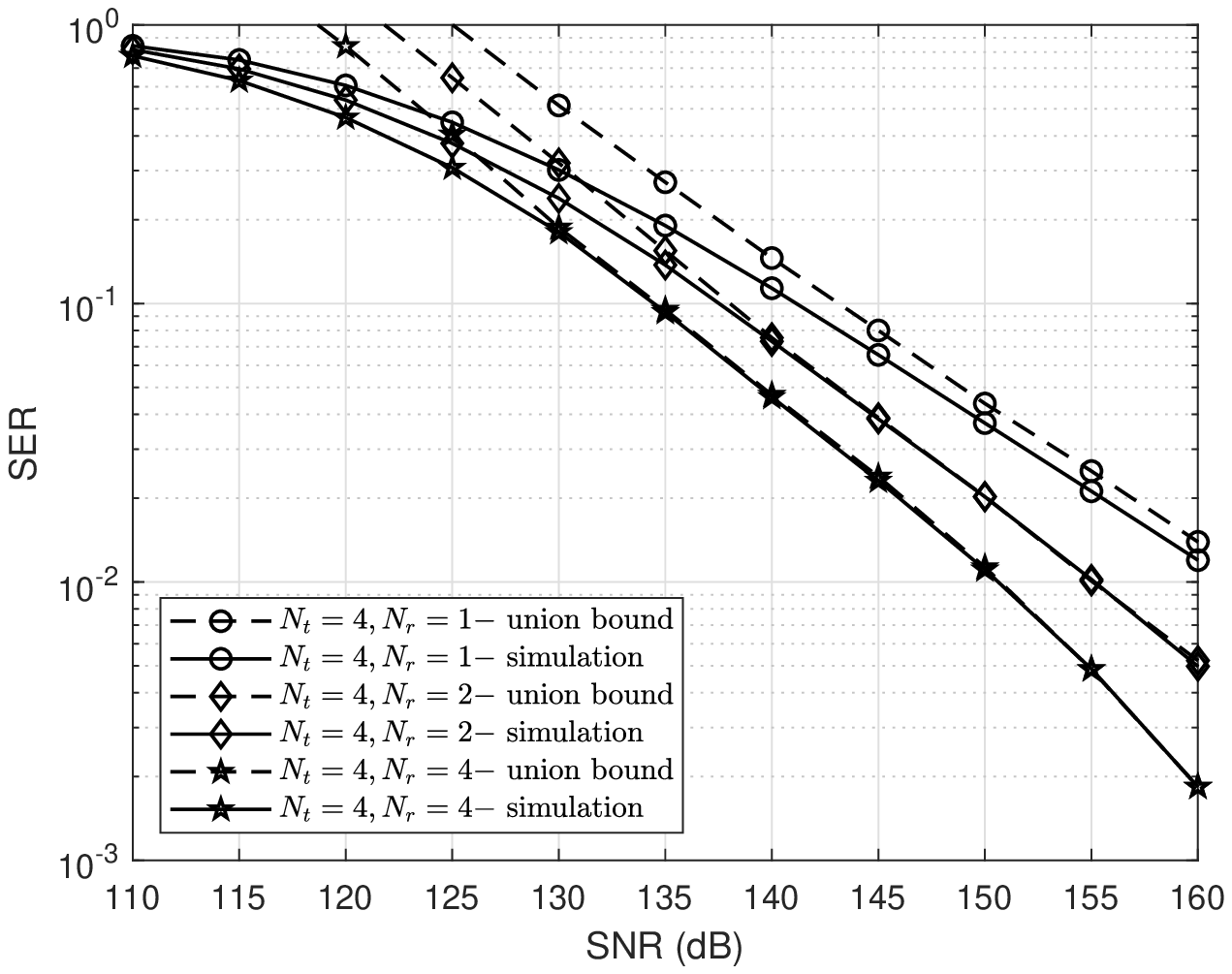}
  \caption{CER plots for 4 Tx antennas and varying number of Rx antennas with union bound for $\gamma$=0.25.}
  \label{SER_1}
\end{minipage}%
\begin{minipage}{.5\textwidth}
  \centering
  \includegraphics[width=.8\linewidth]{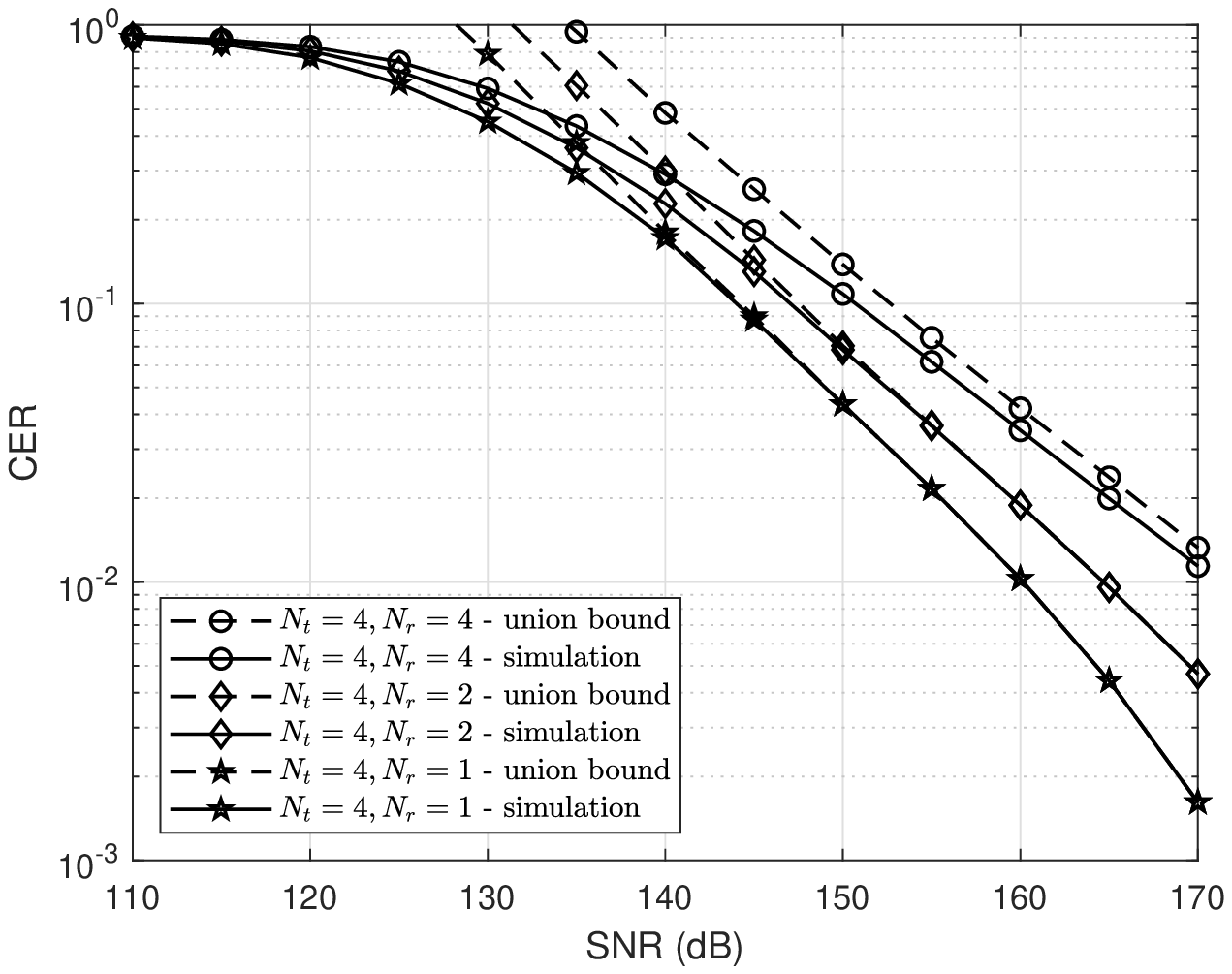}
  \caption{CER plots for 4 Tx antennas and varying number of Rx antennas with union bound for $\gamma$=0.75.}
  \label{SER_2}
\end{minipage}
\end{figure}

Assuming all LEDs are identical and all PDs are identical, the parameters \cite{10A} for simulation  are as follows:\\
Vertical distance from LEDs to surface, L is 2.15 m, cell radius, $r_e$ is 3.55 m, LED semi angle $\Phi_{1/2}$ and PD FOV, $\psi_{fov}$ are both $60^0$, PD responsivity, $R_p$ is 0.4 $A/W$, PD detection area, $A$ is 1 $\mbox{cm}^2$, reflective index $\eta$=1.5, optical filter gain, $T$ is 1, distance from the reference point of the Tx to each Tx antenna, $l'$ is 1 m and distance from the reference point of the Rx to each Rx antenna, $l$ is 5 cm. Since Tx antenna positions are fixed, from the above considerations, their locations on co-ordinate axes in meters can be given as (1,0), (0,1), (-1,0) and (0,-1) for the Tx antennas $T_1$, $T_2$, $T_3$ and $T_4$ respectively. For $N_t$=5, $T_0$ is considered to be the $5^{th}$ Tx antenna whose location is the reference point itself i.e., (0,0). $r$, $\theta$ are generated from the uniform distributions, $U[0,\,r_e]$ and $U[0,\,2\pi)$, respectively. $R_0=(c_0,\,d_0)=(r\mathrm{cos}\theta,r\mathrm{sin}\theta)$ acts as the reference point for the Rx. Since distance between the reference point of the Rx to other Rx antennas is $l$ and assumed to be 5 cm ie., 0.05 m. Then $R_1=(c_1,\,d_1)=(c_0+0.05,d_0)$, $R_2=(c_2,\,d_2)=(c_0,d_0+0.05)$, $R_3=(c_3,\,d_3)=(c_0-0.05,d_0)$ and $R_4=(c_4,\,d_4)=(c_0,d_0-0.05)$. For 5 Rx antennas case, another Rx antenna is considered at $R_0$. Similar Tx and Rx structures can be designed for any $N_t$ and $N_r$. 
Here we will assume that the Rx has the knowledge of the $\gamma$ used at the Tx.
For the proposed codes, CER simulation results are shown for ML, MMSE, and ZF detectors. In ML equalization, we minimize the Euclidean distance over all possible code matrices and the minimization problem is given as
\begin{equation*}
\hat{\mathbf{\underline{X}}}^a_{ML}=||\mathbf{\underline{Y}}-\sqrt{E_s}\mathbf{\underline{H}}\,\mathbf{\underline{X}}^a||^2,
\end{equation*}
In numerical results, to compare the CER and mutual information for different values of $\gamma$, any constellation used in this paper is properly normalized such that the average energy transmitted per code matrix, $\gamma E_s {N_t}^2$ remains same. This can be done by tuning $E_s$. In the plots SNR in dB is equal to $\gamma E_s {N_t}^2 /N_0$.
In Fig.\ref{SER_1}, Fig.\ref{SER_2} the CER plots using simulation along with union bound are shown for 4 Tx antennas-1 Rx antenna, 4 Tx antennas-2 Rx antennas and 4 Tx antennas-4 Rx antennas cases for $\gamma=0.25$ and $\gamma=0.75$, respectively using ML equalization. It can be seen that as $\gamma$ increases CER performance decreases for same number of Tx antennas and same number of Rx antennas cases. In MMSE equalization, the estimate of $\mathbf{\underline{Y}}$ denoted by $\hat{\mathbf{\underline{X}}}^a_{MMSE}$ is given as
\begin{equation*}
\hat{\mathbf{\underline{X}}}^a_{MMSE}=\left(\mathbf{ \underline{H}}^T+\frac{N_0}{2}\mathbf{\underline{I}} \right)^{-1}\mathbf{\underline{H}}^T \mathbf{\underline{Y}},
\end{equation*}
and for ZF equalization, the estimate of $\mathbf{\underline{Y}}$ denoted by $\hat{\mathbf{\underline{X}}}_{ZF}$ is given as
\begin{equation*}
\hat{\mathbf{\underline{X}}}^a_{ZF}=\mathbf{\underline{H}}^{-1}\mathbf{\underline{Y}},
\end{equation*}

Since $1/N_t \leq \gamma \leq (N_t-1)/N_t$, to decode any received vector after ZF and MMSE equalization, take $f$ number of higher magnitude values in each row of the matrix to be 1 and remaining to be 0 if the $\gamma$ used at the transmitter is $f/N_t$, where, $f\epsilon \left\{ 1,2, \ldots, N_t-1  \right\}$. At this point, if $\hat{\mathbf{\underline{X}}}^a_{ZF}$ or $\hat{\mathbf{\underline{X}}}^a_{MMSE}$ is not a valid code matrix, we randomly choose a message vector as decoded vector.

\textbf{Remark 1:} Since the proposed algorithm is systematic, with ZF and MMSE detectors we need not search exhaustively among the code matrices at the encoder and decoder.

\begin{figure}
\centering
\begin{minipage}{.5\textwidth}
  \centering
  \includegraphics[width=.8\linewidth]{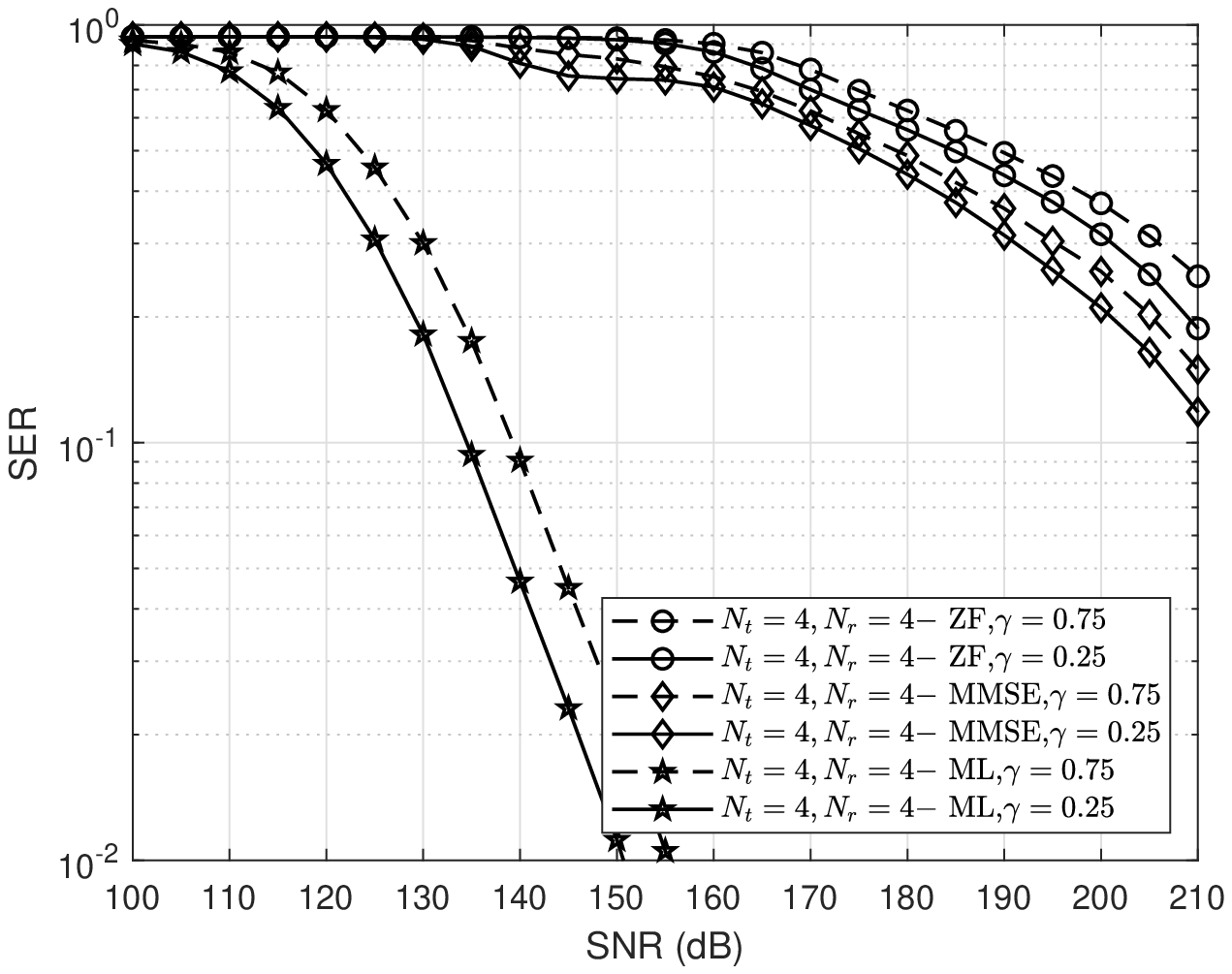}
  \caption{CER plots for 4 Tx antennas and 4 Rx antennas with ZF, MMSE and ML detectors.}
  \label{SER_3}
\end{minipage}%
\begin{minipage}{.5\textwidth}
  \centering
  \includegraphics[width=.8\linewidth]{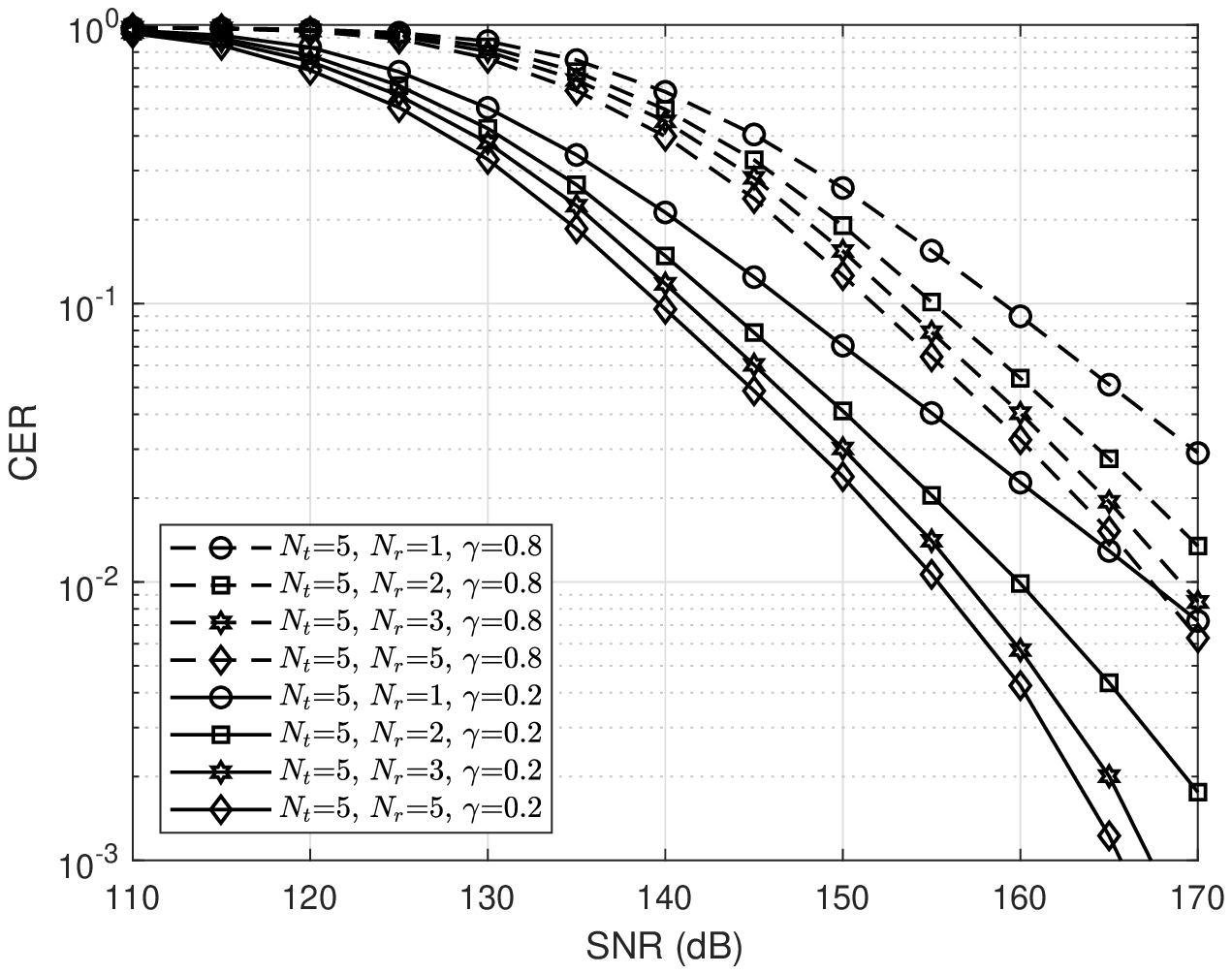}
  \caption{CER plots for 5 Tx antennas and varying number of Rx antennas.}
  \label{SER_4}
\end{minipage}
\end{figure}

\begin{figure}
\centering
\begin{minipage}{.5\textwidth}
  \centering
  \includegraphics[width=.9\linewidth]{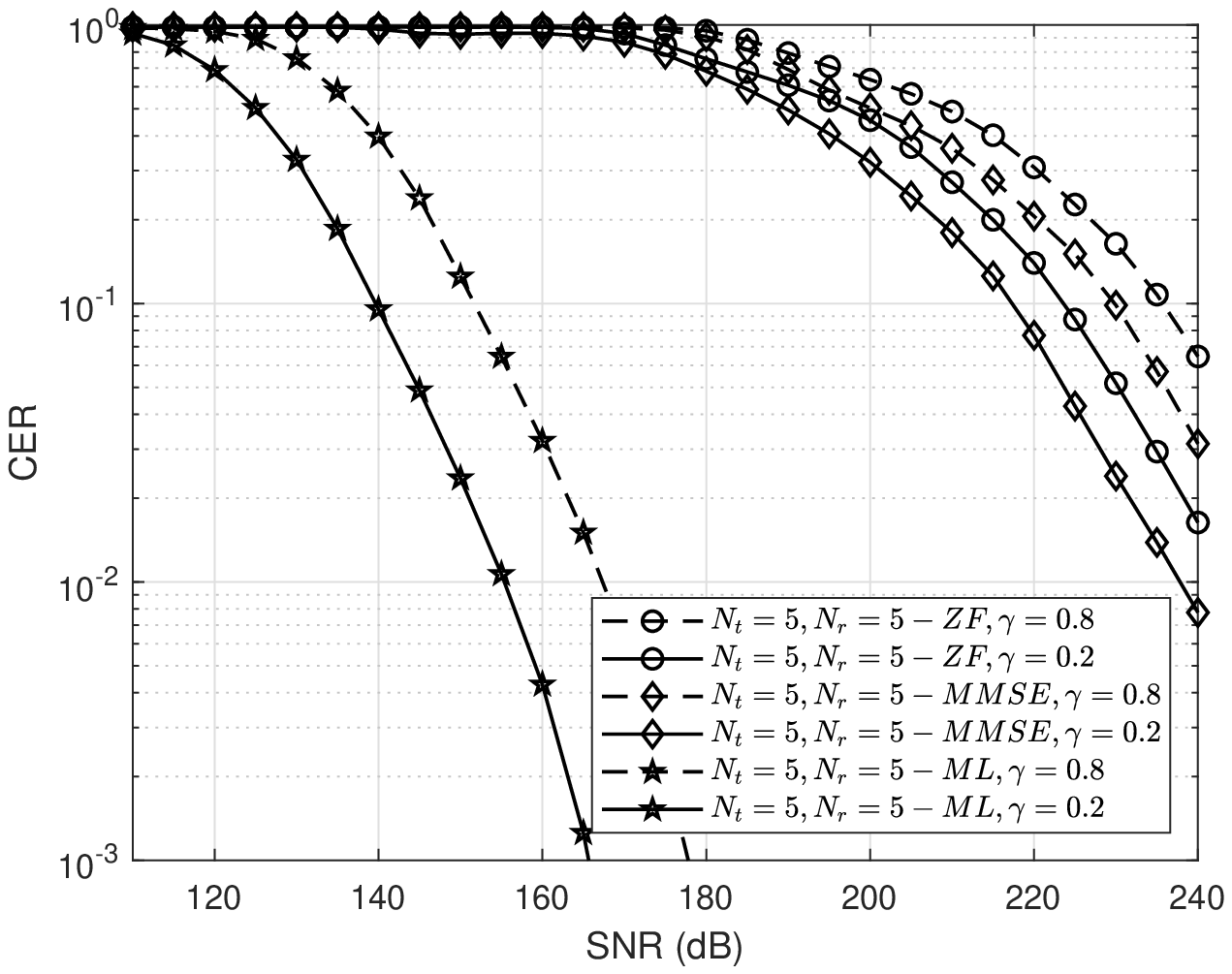}
  \caption{CER plots for 5 Tx antennas and 5 Rx antennas with ZF, MMSE and ML detectors.}
  \label{SER_5}
\end{minipage}%
\begin{minipage}{.5\textwidth}
  \centering
  \includegraphics[width=.85\linewidth]{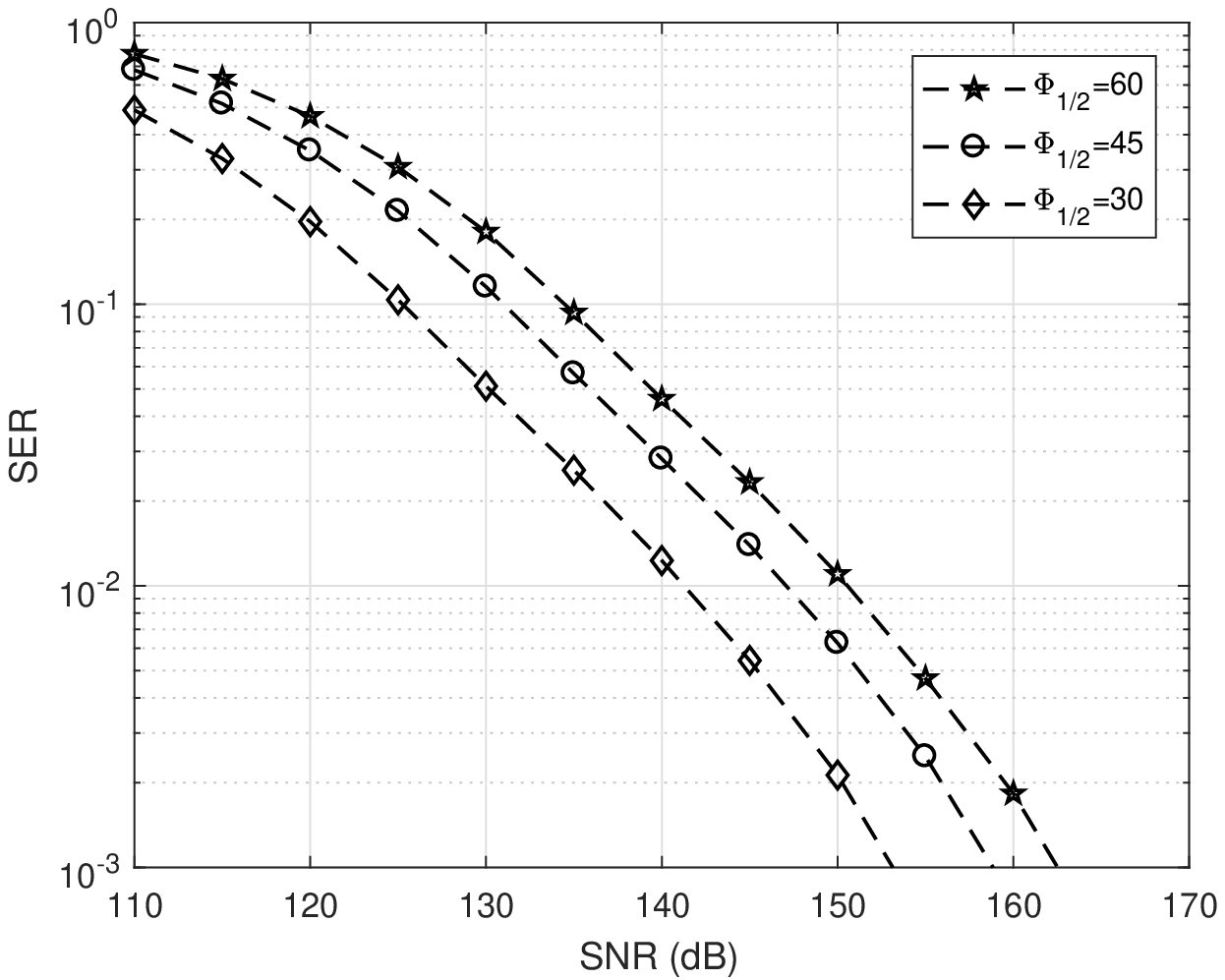}
  \caption{CER plots for 4 Tx antennas and 4 Rx antennas for three values of semi angle of LED (in degrees), $\Phi_{1/2}$ for  $\gamma$=0.25.}
  \label{SER_semiangle}
\end{minipage}
\end{figure}

In Fig.\ref{SER_3} and Fig.\ref{SER_5}, CER plots are shown for ZF, MMSE and ML detectors for 4 Tx antennas and 4 Rx antennas case for $\gamma$=0.25, $\gamma$=0.75 and  for 5 Tx antennas and 5 Rx antennas case for $\gamma$=0.20, $\gamma$=0.80 respectively. In Fig.\ref{SER_4}, the CERs for 5 Tx antennas and varying number of Rx antennas case for $\gamma=0.2$ and $0.8$ are compared where the effect of dimming and number of receive antennas on the system performance is clearly seen.
In Fig.\ref{MI_1}, the mutual information plots for 4 Tx antennas and 1, 2, 3 and 4 Rx antennas are shown. Similarly in Fig.\ref{MI_2}, the mutual information plots for 5 Tx antennas and 1, 2, 3, 4 and 5 Rx antennas are shown. It can be seen that, the mutual information increases as the number of Rx antennas increase before it saturates to the code rate. The difference in the code rates achieved is clearly seen from the saturation levels in the mutual information plots. The code rates achieved by 4 Tx antennas and 5 Tx antennas cases are 1.0 bit/slot and 1.2 bits/slot, respectively. In Fig.\ref{SER_semiangle}, the effect of $\Phi_{1/2}$ on CER performance is shown. From \eqref{vlc_channel}, it can be observed that channel gain increases as $\Phi_{1/2}$ decreases and hence there is an improvement in CER performance with reduction in $\Phi_{1/2}$. 


%

%
\ifCLASSOPTIONcaptionsoff
  \newpage
\fi
Some concluding remarks and future work plan are discussed in the next section.
\section{Conclusions}
The proposed codes with generalized algorithms for encoding, decoding and to achieve desired dimming are presented. The rates achieved, run-lengths for least dimming level achieved conditioned on MFTP are discussed. CER plots for different number of transmit and receive antennas with different dimming levels are shown along with  the mathematical expressions for union bound. The plots for mutual information are also presented along with the expressions. We have also discussed minimum Hamming distance for the proposed codes. In future, we plan to test the performance of the proposed codes on hardware.
\section*{Appendix A}
The Lemma 1 is proved by assuming two numbers $w$ and $w+z$, $z\neq$0 such that both the numbers lie between 0 and $2^{N_t}-1$.
Let $R_1=w$ such that $0 \leq R_1 \leq 2^{N_t}-1$. Then,

 $P_1^w=\left \lfloor\frac{R_1}{(N_t-1)!}\right \rfloor $,
 
$P_2^w=\left \lfloor\frac{w \mathrm{mod} (N_t-1)!}{(N_t-2)!}\right \rfloor $,

 $P_3^w=\left \lfloor\frac{\left(w \mathrm{mod} (N_t-1)!\right)\mathrm{mod}(N_t-2)! }{(N_t-3)!}\right \rfloor $.
  
Similarly, $(N_t-2)^{th}$ term is given as
 
$P_{N_t-2}^k=\left \lfloor\frac{\left(\left(w \mathrm{mod} (N_t-1)!\right)\mathrm{mod}(N_t-2)!\ldots \right)3! }{2!}\right \rfloor $. 

Similarly, when $R_1=w+z$ and $z \neq$0. $P_i$s are given as  

$P_1^{w+z}=\left \lfloor\frac{R_1}{(N_t-1)!}\right \rfloor $,

$P_2^{w+z}=\left \lfloor\frac{{(w+z)} \mathrm{mod} (N_t-1)!}{(N_t-2)!}\right \rfloor $, 

$P_3^{w+z}=\left \lfloor\frac{\left({(w+z)} \mathrm{mod} (N_t-1)!\right)\mathrm{mod}(N_t-2)! }{(N_t-3)!}\right \rfloor $.

Similarly, $(N_t-2)^{th}$ term is given as
 
$P_{N_t-2}^{w+z}=\left \lfloor\frac{\left(\left({(w+z)} \mathrm{mod} (N_t-1)!\right)\mathrm{mod}(N_t-2)!\ldots \right)3! }{2!}\right \rfloor $.

Let us assume the $P_i$s are same for both $w$ and $w+z$ till $i=N_t-3$.
For the $(N_t-2)^{th}$ value of $P_i$, it is seen that the modulo operation in the numerator is done with 3! at the end. So the numerator value at this stage lies among \{0, 1, 2, 3, 4, 5\}. This value is divided by 2! and floored to get $P_{N_t-2}$ which can be \{0, 0, 1, 1, 2, 2\} respectively. If $w$ and $w+z$ leave different $P_i$ at this place, then the Lemma 1 statement is satisfied. If $P_{N_t-2}$ is same for both $w$ and $w+z$ then they differ in the $(N_t-1)^{th}$ and $N_t^{th}$ $P_i$s because of the distinct $R_i$s. In brief, at $(N_t-2)^{th}$ $P_i$, the $R_i$ is found using modulo with 3!. If the two numbers, $w$ and $w+z$ are such that they give different remainders when divided by 6, they differ in $P_i$s at that step itself otherwise in the later steps. Similar explanation holds true for any $P_{N_t-i}$ and hence \{$P_i$\} has at least one element not in common for $z \neq$0. This concludes the proof of Lemma 1. $\hfill \rule{1.5ex}{1.5ex}$

\section*{Appendix B}
The code matrices for mapping 0 to 15 for $\gamma$=0.25 and $N_t$=4 are given below
\begin{tiny}
\[
\begin{bmatrix}
     0  &   0  &   0 &    1\\
     0  &   0  &   1 &    0\\
     0  &   1  &   0 &    0\\
     1  &   0  &   0 &    0\\    
\end{bmatrix}
,
\begin{bmatrix}
     0  &   0  &   0 &    1\\
     0  &   0  &   1 &    0\\
     1   &  0   &  0   &  0\\
     0   &  1  &   0    & 0\\
\end{bmatrix}
,
\begin{bmatrix}
     0    & 0&     0&     1\\
     0   &  1    & 0   &  0\\
     0   &  0   &  1    & 0\\
     1   &  0  &   0    & 0\\
\end{bmatrix} 
,
\begin{bmatrix}
     0 &    0    & 0 &    1\\
     0  &   1    & 0  &   0\\
     1  &   0   &  0   &  0\\
     0  &   0  &   1   &  0\\
\end{bmatrix}  
,\]
\end{tiny}
\begin{tiny}
\[
\begin{bmatrix}
     0  &   0 &    0 &   1\\
     1  &   0 &    0 &    0\\
     0  &   0 &    1 &    0\\
     0  &   1 &    0 &    0  \\   
\end{bmatrix}
,
\begin{bmatrix}
     0 &    0  &   0 &    1\\
     1 &    0  &   0 &    0\\
     0 &    1  &   0 &    0\\
     0 &    0  &   1 &    0\\
\end{bmatrix}
, 
\begin{bmatrix}
     0 &    0 &    1 &    0\\
     0 &    0 &    0 &    1\\
     0 &    1 &    0 &    0\\
     1 &    0 &    0 &    0\\
\end{bmatrix} 
,
\begin{bmatrix}
     0 &    0  &   1 &    0\\
     0 &    0   &  0 &    1\\
     1 &    0  &   0 &    0\\
     0 &    1 &    0 &    0\\
\end{bmatrix}  
,\]
\end{tiny}
\begin{tiny}
\[
\begin{bmatrix}
     0 &    0 &    1 &    0\\
     0 &    1 &    0 &    0\\
     0 &    0 &    0 &    1\\
     1 &    0 &    0 &    0  \\    
\end{bmatrix}
,
\begin{bmatrix}
     0 &    0 &    1  &   0\\
     0 &    1 &    0  &   0\\
     1 &    0 &    0  &   0\\
     0 &    0 &    0  &   1\\
\end{bmatrix}
,
\begin{bmatrix}
     0 &    0  &   1  &   0\\
     1 &    0  &   0  &   0\\
     0 &    0  &   0  &   1\\
     0 &    1  &   0  &   0\\
\end{bmatrix}
, 
\begin{bmatrix}
     0  &   0  &   1  &   0\\
     1  &   0  &   0  &   0\\
     0  &   1  &   0  &   0\\
     0  &   0  &   0  &   1\\
\end{bmatrix}  
,\]
\end{tiny}
\begin{tiny}
\[
\begin{bmatrix}
     0  &   1 &    0  &   0\\
     0  &   0 &    0  &   1\\
     0  &   0 &    1  &   0\\
     1  &   0 &    0  &   0  \\ 
\end{bmatrix}
, 
\begin{bmatrix}
     0 &    1  &   0 &    0\\
     0 &    0  &   0 &    1\\
     1 &    0  &   0 &    0\\
     0 &    0  &   1 &    0\\
\end{bmatrix}
, 
\begin{bmatrix}
     0  &   1 &    0  &   0\\
     0  &   0 &    1  &   0\\
     0  &   0 &    0  &   1\\
     1  &   0 &    0  &   0\\
\end{bmatrix} 
, 
\begin{bmatrix}
     0 &    1 &    0  &   0\\
     0 &    0 &    1  &   0\\
     1 &    0 &    0  &   0\\
     0 &    0 &    0  &   1\\
\end{bmatrix}  
.\]
\end{tiny}

The code matrices for mapping 0 to 15 for $\gamma$=0.50 and $N_t$=4 are given below
\begin{tiny}
\[
\begin{bmatrix}
     1  &   0  &   0 &    1\\
     0  &   0  &   1 &    1\\
     0  &   1  &   1 &    0\\
     1  &   1  &   0 &    0\\    
\end{bmatrix}
,
\begin{bmatrix}
     1  &   0  &   0 &    1\\
     0  &   0  &   1 &    1\\
     1   &  1   &  0   &  0\\
     0   &  1  &   1    & 0\\
\end{bmatrix}
,
\begin{bmatrix}
     1    & 0&     0&     1\\
     0   &  1    & 1   &  0\\
     0   &  0   &  1    & 1\\
     1   &  1  &   0    & 0\\
\end{bmatrix} 
,
\begin{bmatrix}
     1 &    0    & 0 &    1\\
     0  &   1    & 1  &   0\\
     1  &   1   &  0   &  0\\
     0  &   0  &   1   &  1\\
\end{bmatrix}  
,\]
\end{tiny}
\begin{tiny}
\[
\begin{bmatrix}
     1  &   0 &    0 &   1\\
     1  &   1 &    0 &    0\\
     0  &   0 &    1 &    1\\
     0  &   1 &    1 &    0  \\   
\end{bmatrix}
,
\begin{bmatrix}
     1 &    0  &   0 &    1\\
     1 &    1  &   0 &    0\\
     0 &    1  &   1 &    0\\
     0 &    0  &   1 &    1\\
\end{bmatrix}
, 
\begin{bmatrix}
     0 &    0 &    1 &    1\\
     1 &    0 &    0 &    1\\
     0 &    1 &    1 &    0\\
     1 &    1 &    0 &    0\\
\end{bmatrix} 
,
\begin{bmatrix}
     0 &    0  &   1 &    1\\
     1 &    0   &  0 &    1\\
     1 &    1  &   0 &    0\\
     0 &    1 &    1 &    0\\
\end{bmatrix}  
,\]
\end{tiny}
\begin{tiny}
\[
\begin{bmatrix}
     0 &    0 &    1 &    1\\
     0 &    1 &    1 &    0\\
     1 &    0 &    0 &    1\\
     1 &    1 &    0 &    0  \\    
\end{bmatrix}
,
\begin{bmatrix}
     0 &    0 &    1  &   1\\
     0 &    1 &    1  &   0\\
     1 &    1 &    0  &   0\\
     1 &    0 &    0  &   1\\
\end{bmatrix}
,
\begin{bmatrix}
     0 &    0  &   1  &   1\\
     1 &    1  &   0  &   0\\
     1 &    0  &   0  &   1\\
     0 &    1  &   1  &   0\\
\end{bmatrix}
, 
\begin{bmatrix}
     0  &   0  &   1  &   1\\
     1  &   1  &   0  &   0\\
     0  &   1  &   1  &   0\\
     1  &   0  &   0  &   1\\
\end{bmatrix}  
,\]
\end{tiny}
\begin{tiny}
\[
\begin{bmatrix}
     0  &   1 &    1  &   0\\
     1  &   0 &    0  &   1\\
     0  &   0 &    1  &   1\\
     1  &   1 &    0  &   0  \\ 
\end{bmatrix}
, 
\begin{bmatrix}
     0 &    1  &   1 &    0\\
     1 &    0  &   0 &    1\\
     1 &    1  &   0 &    0\\
     0 &    0  &   1 &    1\\
\end{bmatrix}
, 
\begin{bmatrix}
     0  &   1 &    1  &   0\\
     0  &   0 &    1  &   1\\
     1  &   0 &    0  &   1\\
     1  &   1 &    0  &   0\\
\end{bmatrix} 
, 
\begin{bmatrix}
     0 &    1 &    1  &   0\\
     0 &    0 &    1  &   1\\
     1 &    1 &    0  &   0\\
     1 &    0 &    0  &   1\\
\end{bmatrix}  
.\]
\end{tiny}
The code matrices for mapping 0 to 15 for $\gamma$=0.75 and $N_t$=4 are given below
\begin{tiny}
\[
\begin{bmatrix}
     1  &   1  &   0 &    1\\
     1  &   0  &   1 &    1\\
     0  &   1  &   1 &    1\\
     1  &   1  &   1 &    0\\    
\end{bmatrix}
,
\begin{bmatrix}
     1  &   1  &   0 &    1\\
     1  &   0  &   1 &    1\\
     1   &  1   &  1   &  0\\
     0   &  1  &   1    & 1\\
\end{bmatrix}
,
\begin{bmatrix}
     1    & 1&     0&     1\\
     0   &  1    & 1   &  1\\
     1   &  0   &  1    & 1\\
     1   &  1  &   1    & 0\\
\end{bmatrix} 
,
\begin{bmatrix}
     1 &    1    & 0 &    1\\
     0  &   1    & 1  &   1\\
     1  &   1   &  1   &  0\\
     1  &   0  &   1   &  1\\
\end{bmatrix}  
,\]
\end{tiny}
\begin{tiny}
\[
\begin{bmatrix}
     1  &   1 &    0 &   1\\
     1  &   1 &    1 &    0\\
     1  &   0 &    1 &    1\\
     0  &   1 &    1 &    1  \\   
\end{bmatrix}
,
\begin{bmatrix}
     1 &    1  &   0 &    1\\
     1 &    1  &   1 &    0\\
     0 &    1  &   1 &    1\\
     1 &    0  &   1 &    1\\
\end{bmatrix}
, 
\begin{bmatrix}
     1 &    0 &    1 &    1\\
     1 &    1 &    0 &    1\\
     0 &    1 &    1 &    1\\
     1 &    1 &    1 &    0\\
\end{bmatrix} 
,
\begin{bmatrix}
     1 &    0  &   1 &    1\\
     1 &    1   &  0 &    1\\
     1 &    1  &   1 &    0\\
     0 &    1 &    1 &    1\\
\end{bmatrix}  
,\]
\end{tiny}
\begin{tiny}
\[
\begin{bmatrix}
     1 &    0 &    1 &    1\\
     0 &    1 &    1 &    1\\
     1 &    1 &    0 &    1\\
     1 &    1 &    1 &    0  \\    
\end{bmatrix}
,
\begin{bmatrix}
     1 &    0 &    1  &   1\\
     0 &    1 &    1  &   1\\
     1 &    1 &    1  &   0\\
     1 &    1 &    0  &   1\\
\end{bmatrix}
,
\begin{bmatrix}
     1 &    0  &   1  &   1\\
     1 &    1  &   1  &   0\\
     1 &    1  &   0  &   1\\
     0 &    1  &   1  &   1\\
\end{bmatrix}
, 
\begin{bmatrix}
     1  &   0  &   1  &   1\\
     1  &   1  &   1  &   0\\
     0  &   1  &   1  &   1\\
     1  &   1  &   0  &   1\\
\end{bmatrix}  
,\]
\end{tiny}
\begin{tiny}
\[
\begin{bmatrix}
     0  &   1 &    1  &   1\\
     1  &   1 &    0  &   1\\
     1  &   0 &    1  &   1\\
     1  &   1 &    1  &   0  \\ 
\end{bmatrix}
, 
\begin{bmatrix}
     0 &    1  &   1 &    1\\
     1 &    1  &   0 &    1\\
     1 &    1  &   1 &    0\\
     1 &    0  &   1 &    1\\
\end{bmatrix}
, 
\begin{bmatrix}
     0  &   1 &    1  &   1\\
     1  &   0 &    1  &   1\\
     1  &   1 &    0  &   1\\
     1  &   1 &    1  &   0\\
\end{bmatrix} 
, 
\begin{bmatrix}
     0 &    1 &    1  &   1\\
     1 &    0 &    1  &   1\\
     1 &    1 &    1  &   0\\
     1 &    1 &    0  &   1\\
\end{bmatrix}  
.\]
\end{tiny}
\section*{Acknowledgment}
This work was supported in part by the Department of Science and Technology (DST), Govt. of India (Ref. No. TMD/CERI/BEE/2016/059(G)).

\end{document}